\documentclass[12pt]{article}
\usepackage{amsmath,amsthm,amssymb,mathtools}
\usepackage{graphicx}
\usepackage{enumerate}
\usepackage{natbib}
\usepackage{caption}
\usepackage{url} % not crucial - just used below for the URL 
\usepackage[final, bookmarks, bookmarksnumbered=true, bookmarksopen=true, colorlinks=true, hyperfootnotes, linkcolor=black, anchorcolor=blue, citecolor=blue, filecolor=blue, menucolor=blue, runcolor=blue, urlcolor=black]{hyperref}
\usepackage{cleveref}
\usepackage{tikz}
\usepackage{booktabs}
\usepackage{threeparttable}
\usepackage{bbm}
\usepackage{comment}
\usepackage{changepage}
\usepackage{pgffor}
\usetikzlibrary{automata,positioning,fit, shapes.geometric,calc}

\newcommand{\blind}{1}

\newtheorem{theorem}{Theorem}
\newtheorem{lemma}{Lemma}
\newtheorem{assumption}{Assumption}
\newtheorem{corollary}{Corollary}
\newtheorem{proposition}{Proposition}

\newcommand\blfootnote[1]{
    \begingroup
    \renewcommand\thefootnote{}\footnote{#1}
    \addtocounter{footnote}{-1}
    \endgroup
}

\begin{document}

\def\spacingset#1{\renewcommand{\baselinestretch}%
{#1}\small\normalsize} \spacingset{1}

%%%%%%%%%%%%%%%%%%%%%%%%%%%%%%%%%%%%%%%%%%%%%%%%%%%%%%%%%%%%%%%%%%%%%%%%%%%%%%

\if1\blind
{
  \title{\bf Instrumental variables with treatment clusters}
  \author{Didier Nibbering\thanks{Department of Econometrics and Business Statistics, Monash University, Melbourne, Australia. didier.nibbering@monash.edu} \and
Matthijs Oosterveen\thanks{Department of Economics, Lisbon School of Economics and Management, and Advance/ISEG Research, University of Lisbon, Lisbon, Portugal. oosterveen@iseg.ulisboa.pt}  \and
Pedro Luis Silva\thanks{Department of Applied Economics, Universitat Aut\`{o}noma de Barcelona, Barcelona, Spain, and Centre for Research in Higher Education Policies, Porto, Portugal. pedroluis.silva@uab.cat} 
\blfootnote{
We thank Sacha Becker, David Frazier, Sacha Kapoor, Joppe de Ree, Serguei Maliar, Denni Tommasi, and Dinand Webbink for valuable comments. The paper has also benefited from the comments of participants at LESE 2022, PEJ 2022, IPDC2024, 2025 Labour Econometrics Workshop, the Nova SBE Economics of Education seminar, and the Erasmus University Rotterdam Brown Bag seminar. 
We thank the Dire\c{c}\~{a}o-Geral do Ensino Superior (DGES) and the Dire\c{c}\~{a}o-Geral de Estat\'{i}sticas da Educa\c{c}\~{a}o e Ci\^{e}ncia (DGEEC) for the access to the two data sets used in this paper. The data sets were merged at the premises of DGEEC and cannot be disclosed to anyone who has not signed a confidentiality agreement with DGEEC.
Oosterveen and Silva acknowledge financial support from FCT, Fundação para a Ciência e a Tecnologia (Portugal). Oosterveen acknowledges support from grants PTDC/EGE-OGE/28603/2017 and UID/06522/2025, and Silva from grants 2023.08911.CEECIND/CP2880/CT0001 and UID/00757/2025. Silva also acknowledges financial support from MICIU, Ministerio de Ciencia, Innovaci\'{o}n y Universidades (Spain), through grant PID2024-162349NB-I00.
The authors have no relevant or material financial
interests that relate to the research described in this paper. All omissions and errors are our own.
}}
  \maketitle
} \fi

\if0\blind
{
  \bigskip
  \bigskip
  \bigskip
  \begin{center}
    {\LARGE\bf Instrumental variables with treatment clusters}
\end{center}
  \medskip
} \fi

\bigskip
\begin{abstract}
Many policy evaluations include multiple unordered treatments. A common approach is to aggregate these treatments into clusters and apply instrumental variable (IV) methods at the treatment cluster level: The clusters are treated as binary treatments and the IV assumptions are invoked at the cluster level. This paper shows that IV with treatment clusters identifies a weighted average of local average treatment effects (LATEs) under a set of assumptions at the unclustered treatment level. When data on unclustered treatment choice is available, we propose testable necessary conditions for the causal interpretation of IV with treatment clusters. We illustrate our methods with empirical applications to the causal effects of fields of study aggregated into STEM and non-STEM clusters or eight broad study directions.
\end{abstract}

\noindent%
\vfill

\newpage
\spacingset{1.65} % DON'T change the spacing!
\section{Introduction}\label{sec:intro}

% introduce with example STEM
Many policy evaluations involve multiple unordered treatments, such as location decisions by firms or families \citep{chetty2016effects} or student fields of study choices \citep{altonji2016analysis}. Researchers often aggregate these treatments into clusters. For instance, STEM and non-STEM programs \citep{ng2024returns}, Business and non-Business majors \citep{andrews2017risky}, or broad study directions \citep{dahl2023high}, all consist of fields of study such as Physics or Finance. When treatment choice is endogenous, instrumental variable (IV) methods may be used to estimate a local average treatment effect (LATE) of a cluster of treatments. A STEM degree is then considered as a binary treatment choice, and the IV assumptions are invoked at the cluster level. However, individuals choose among treatments within the clusters, and therefore the instruments may induce changes in treatment take up across and within clusters. Hence, treatments have to be considered at the unclustered level as otherwise violations of the IV assumptions may stay undetected \citep{imbens2015causal}. This raises the question of what parameter IV estimation at the cluster level is targeting, and under what conditions this parameter has a causal interpretation. 

% what we do
This paper studies IV estimation with treatments aggregated into clusters in a setting where both treatment and instrument are multi-valued unordered variables defined over the same support. We show that IV with clustered treatments identifies a weighted average of LATEs under a set of assumptions imposed at the unclustered treatment level. This weighted average includes the LATEs of all pairs of treatments across two clusters, and all instrument values associated with these clusters. The assumptions clarify which potential treatment choices have to be excluded, which does not directly follow from the IV assumptions of \citet{imbens1994identification} invoked at the cluster level. The assumptions are less strict than the IV assumptions required for identifying LATEs between the unclustered treatments \citep{bhuller20242sls}. We show that this result applies to settings with an arbitrary number of treatment clusters and a corresponding number of instruments. We also derive necessary conditions for a causal interpretation of the IV estimand with treatment clusters that are testable using data at the unclustered treatment level.

% relevance
Treatment evaluation often considers clusters of treatments instead of the actual treatments for three reasons. First, administrative data on treatments with small enrollment numbers may not be disclosed due to privacy concerns. Second, the estimation of a large number of pairwise treatment effects may suffer from a lack of accuracy. Third, a large number of estimates may be less relevant if the policy is designed at a less fine-grained level. Our results are applicable to these settings, and help practitioners interpret the parameter targeted by an IV analysis at the cluster level, and justify the IV assumptions required for a causal interpretation of this parameter. 

% discuss application and results.
To illustrate the empirical relevance of our methods, we study the causal effect of clusters of fields of study on the number of study credit points. We have access to an administrative data set that includes four cohorts in the Portuguese public higher education system, from which we exploit a natural experiment to construct instruments in a fuzzy regression discontinuity design (RDD). We estimate treatment cluster effects using the STEM versus non-STEM classification and eight broad academic clusters defined according to the International Standard Classification of Education (ISCED).

% discuss related literature with treatment clusters
To the best of our knowledge, we are the first to analyze the causal interpretation of the IV estimand with clusters of multiple unordered treatments. In a related setting, but with an ordered multivalued treatment, \citet{angrist1995two,marshall2016coarsening,andresen2021instrument} discuss the implications of binarising the multivalued treatment. Treatment clusters have also been discussed in combination with conditional independence assumptions. For instance, \citet{lechner2002program} discusses how to estimate the effect of treatment clusters in this setting. \citet{heiler2022effect} decompose treatment effect heterogeneity with two treatment clusters into heterogeneous treatment responses and heterogeneity in the actual treatment take up.

There is an extensive literature that aims to identify pairwise unclustered treatment effects in the presence of multiple unordered treatments. When using multiple instrumental variables, \cite{behaghel2013robustness,kirkeboen2016field,dahl2023high,bhuller20242sls,heiensen2025instrumental} allow each instrument to induce the choice of only one particular treatment, which is more restrictive than the assumptions required for an analysis at the treatment cluster level. \cite{heckman2018unordered} and \cite{pinto2021beyond} discuss alternative assumptions for identifying multiple (local) average treatment effects with multiple discrete instruments. \cite{lee2018identifying} and \cite{mountjoy2021community} discuss this setting with continuous instruments. \cite{blackwell2017instrumental} discusses the identification of interaction effects between multiple treatments with binary instruments. \cite{kline2016evaluating} consider the multiple treatment case with one instrument.

The paper proceeds as follows. \Cref{sec:mut} introduces the setting with IVs for multiple unordered treatments. \Cref{sec:clusters} analyses IV estimation with treatment clusters. \Cref{sec:individual} derives testable necessary conditions for the identifying assumptions, and \Cref{sec:covariates} extends the results to a fuzzy RDD. \Cref{sec:Application} illustrates the empirical relevance of our methods with fields of study as multiple unordered treatments. \Cref{sec:conc} concludes.

\section{IV with multiple unordered treatments}\label{sec:mut}

Suppose we have a multi-valued unordered treatment $D \in \{1,\dots,J\}$ and a multi-valued unordered instrument $Z \in \{1,\dots,J\}$. Define the mutually exclusive binary treatments $D_{j}=\mathbbm{1}[D=j]$ and the mutually exclusive binary instruments $Z_{j}=\mathbbm{1}[Z=j]$. Note that $D=\sum_{j=1}^J j \times D_j$ and $Z=\sum_{j=1}^J j \times Z_j$. This setting, in which treatment and instrument have the same number of values and use the same labels, is common in IV applications with multiple treatments \citep{bhuller20242sls}. \Cref{sec:IVestimand2} discusses how distinct treatment and instrument supports can be accommodated in our treatment cluster framework. Treatments that are not mutually exclusive can always be made into mutually exclusive treatments by defining new categories for each combination of treatments.

Define $Y$ as the observed outcome of interest. Suppose that the effect on $Y$ of taking treatment $j$ compared to treatment $k$ is heterogeneous across individuals. The instrument $Z_j=1$ if the policymaker intends to encourage take up of treatment $j$, but in practice this initial encouragement may also induce individuals to take up other treatments $l\neq j$. We define $D(z)$ as the potential treatment choice when $Z$ is externally set to $z$, with $z=1,\dots,J$. The dummy variable $D_j(z)$ equals one if treatment $j$ is chosen if $Z=z$, and hence $D_j=\sum_{l=1}^J D_j(l)\times Z_l$ for each $j$. Similarly, we define $Y(d,z)$ as the potential outcome when $D=d$ and $Z=z$, with $d=1,\dots,J$ and $z=1,\dots,J$. 
We make the following standard IV assumptions:
\begin{assumption}[IV assumptions]\label{ass:standardIV}\
\renewcommand{\theenumi}{\alph{enumi}}
\begin{enumerate}
    \item (Independence) $Y(d,z),D(z) \perp Z$ for all $d,z$.
    \item (Exclusion) $Y(d,z)=Y(d)$ for all $d,z$. 
\end{enumerate} 
\end{assumption}

%Give examples of multiple unordered treatments.
Analyzing treatment choice at the unclustered treatment level is not always possible. For instance, consider fields of study as the multiple unordered treatments. Obtaining administrative data on study enrollment at the field of study level may be infeasible, in particular due to privacy constraints with the emergence of smaller degrees. With many small degrees it is challenging to obtain accurate estimates for the pairwise treatment effects between the degrees. Moreover, a large number of treatment effect estimates may be less relevant if the policy is designed at a less granular level, such as the promotion of gender equality in STEM enrollment \citep{kahn2017women}. In such cases, practitioners often aggregate treatments into clusters and analyze the causal effects between these treatment clusters. 

\section{IV with clustered treatment data}\label{sec:clusters}

This section analyses IV estimation with treatment clusters. First, it shows under which assumptions the IV estimand with two treatment clusters identifies a causal parameter, and discusses the interpretation of this parameter. Second, it shows that the results extend to the two-stage least squares (TSLS) estimand with an arbitrary number of clusters. 

\subsection{The IV estimand with two treatment clusters}\label{sec:IVestimand2}

Suppose we have two mutually exclusive clusters of treatments $S_1$ and $S_2$. Define $\tilde{D}_1=\sum_{n\in S_1} D_n$, which equals one if a treatment in cluster $S_1$ is chosen and zero otherwise. Similarly, define $\tilde{Z}_1=\sum_{n\in S_1} Z_n$, which equals one if there is an initial encouragement to take up treatment from cluster $S_1$ and zero otherwise.
We refer to $D_n$ and $Z_n$ as the treatment and instrument at the unclustered treatment level, and to $\tilde{D}_1$ and $\tilde{Z}_1$ as these objects at the treatment cluster level. Consider the IV estimand
\begin{align}\label{eq:beta2}
    \beta= \frac{\Delta\mathbb{E}[Y|\tilde{Z}_1]}{\Delta\mathbb{E}[\tilde{D}_1|\tilde{Z}_1]},
\end{align}
where $\Delta \mathbb{E}[A|\tilde{Z}_1]=\mathbb{E}[A|\tilde{Z}_1=1]-\mathbb{E}[A|\tilde{Z}_1=0]$. This estimand can be used to, for instance, identify the returns to STEM education in an RDD where $Y$ equals salary, $\tilde{D}_1=1$ indicates STEM enrollment, and $\tilde{Z}_1=1$ represents a STEM enrollment offer for students with a STEM application score near the admission cutoff. For the IV estimand to exist, we invoke the IV relevance assumption at the cluster level:
\begin{assumption}[Relevance]\label{ass:relevance}
$\Delta\mathbb{E}[\tilde{D}_1|\tilde{Z}_1]>0$.
\end{assumption}
This imposes relevance only at the treatment cluster level. Hence, the instrument may have no effect on the probability of taking treatments at the unclustered level, as long as it has a nonzero effect on the probability of taking treatment from cluster $S_1$.

%- Show diagram with only two clusters
%- explain identification problem
We assume that the estimand in \eqref{eq:beta2} is considered when the effect of taking treatment from cluster $S_1$ compared to taking treatment from cluster $S_2$ is of interest. A key identification challenge is that individuals may respond in various ways to the instrument. To describe this heterogeneity in responses, we follow \citet{heckman2018unordered} and define a response type as a vector of potential treatment choices for $Z$ fixed at each value of its support: $(D(1),D(2),\dots,D(J))$. We refer to all response types who take treatment $l$ if $Z=j$ and treatment $m$ if $Z=k$ as $(D(j)=l,D(k)=m)$. Similarly, $(D(j)\in S_1, D(k)\in S_2)$ describes all types who take treatment from $S_1$ if $Z=j$ and treatment from $S_2$ if $Z=k$. 

\Cref{tab:shifts2} shows four different response types if we both consider the treatments and the instruments at the cluster level. These include all possible types, if we exclude individuals with $D(j)=D(k)$ if $j\in S_1$ and $k \in S_2$, who do not contribute to the instrument-induced treatment contrasts underlying the IV estimand. The following lemma shows that the identification of the effect of taking treatment from cluster $S_1$ compared to taking treatment from $S_2$ requires all response types to satisfy $(D(j)\in S_1, D(k)\in S_2)$ with $j\in S_1$ and $k \in S_2$. To prevent unwanted responses, additional assumptions are required. 

\begin{table}[t!]
    \centering
    \caption{Response types with two treatment clusters}
    \renewcommand{\arraystretch}{0.85}    
    \label{tab:shifts2}
    \begin{tabular}{lll}
    \toprule\toprule
      Response types with $D(j)\neq D(k)$ & Contribution \Cref{lem:decomp}   & Excluded by\\
      \midrule
      $(D(j)\in S_1, D(k)\in S_2)$ & Effects across $S_1$ and $S_2$  & \\
      $(D(j)\in S_2, D(k)\in S_1)$ & Negative effects  & \Cref{ass:mono} \\
      $(D(j)\in S_1, D(k)\in S_1)$ & $B_1$  & \Cref{ass:mono} \\
      $(D(j)\in S_2, D(k)\in S_2)$ & $B_2$  & \Cref{ass:irrelevance} \\
      \bottomrule\bottomrule
    \end{tabular}
    \begin{tablenotes}
    \footnotesize
    \item[] Notes: For $j\in S_1$ and $k\in S_2$, the rows are mutually exclusive and exhaustive among response types satisfying $D(j)\neq D(k)$. 
    \end{tablenotes}
\end{table}

\begin{lemma}[Decomposition IV estimand]\label{lem:decomp}\ \\
Under Assumptions~\ref{ass:standardIV} and \ref{ass:relevance}, it holds for ${\beta}$ in \eqref{eq:beta2} that
\begin{align}
    &{\beta} = 
     \frac{\sum_{j\in S_1} \sum_{k\in S_2}\omega_{jk} \left[ \sum_{l \in S_1}\sum_{m  \in S_2} \Delta_{jklm}P_{jklm} - \Delta_{jkml}P_{jkml} \right]+B_1 +B_2}{\sum_{j\in S_1} \sum_{k\in S_2}\omega_{jk}\left[ \sum_{l \in S_1}\sum_{m  \in S_2} P_{jklm} - P_{jkml}\right]},
\end{align}
with $B_c = \sum_{j\in S_1} \sum_{k\in S_2}\omega_{jk}\sum_{l \in S_c}\sum_{m \neq l \in S_c} \Delta_{jklm}P_{jklm}$ and
$\Delta_{jklm} = \mathbb{E}[Y(l)-Y(m)|D(j)=l,D(k)=m]$, 
$P_{jklm} = \mathbb{P}\left[D(j)=l,D(k)=m\right]$, and
$\omega_{jk}=\mathbb{P}[Z=j|Z\in S_1]\mathbb{P}[Z=k|Z\in S_2]$.
\end{lemma}
The proof is deferred to \Cref{A:proof_decomp}.
%Discuss Lemma.
The decomposition in \Cref{lem:decomp} shows that the IV estimand in \eqref{eq:beta2} is not necessarily a positively weighted average of the treatment effects corresponding to the response types $(D(j)\in S_1, D(k)\in S_2)$ with $j\in S_1$ and $k \in S_2$. Under Assumptions~\ref{ass:standardIV} and \ref{ass:relevance}, the probability of these types equals $P_{jklm}$. However, the contributions of the response types $(D(j)\in S_2, D(k)\in S_1)$ with $P_{jkml}$ enter both the numerator and denominator with a negative sign. Response types $(D(j)\in S_1, D(k)\in S_1)$ and $(D(j)\in S_2, D(k)\in S_2)$ generate the terms $B_1$ and $B_2$, respectively. For example, with $S_1=\{1,2\}$, $S_2=\{3,4\}$, $j=1$, and $k=3$, a response $(D(1)=3,D(3)=1)$ contributes to the negatively weighted term, $(D(1)=1,D(3)=2)$ contributes to $B_1$, and $(D(1)=4,D(3)=3)$ contributes to $B_2$.

%Discuss why cluster monotonicity does not help:
\Cref{lem:decomp} also clarifies how IV assumptions invoked at the cluster level affect the interpretation of $\beta$. Previous research using IV with treatment clusters, such as STEM versus non-STEM \citep{ng2024returns}, imposes monotonicity at the cluster level: $\tilde{D}_1(j)\geq\tilde{D}_1(k)$ with $j\in S_1$ and $k \in S_2$. We refer to this condition as cluster monotonicity, which excludes the responses $(D(j)\in S_2, D(k)\in S_1)$ and therefore removes the negative terms in \Cref{lem:decomp}. It does not, however, rule out the response types in $B_1$ and $B_2$. However, IV identification also relies on an exclusion restriction, which imposes additional restrictions if also invoked at the cluster level: the instrument may only affect $Y$ through $\tilde{D}_1$, which imposes that $\Delta_{jklm}=0$ if $(l,m)\in S_1$ or $(l,m)\in S_2$, and therefore $B_1=B_2=0$. 

Instead of invoking monotonicity directly at the cluster level, we apply the monotonicity assumption in \citet{imbens1994identification} at the treatment level within the treatment cluster setting.
\begin{assumption}[Monotonicity] \label{ass:mono}
$D_l(j)\geq D_l(k)$ for all $j,l \in S_a$ and $k \in S_b$, where $a=1$ and $b=2$ with two clusters.
\end{assumption}
This assumption rules out all individuals for whom $Z \in S_1$ makes it less likely to take up a specific treatment in $S_1$. This excludes the response types $(D(j) \in S_2, D(k) \in S_1)$, which are also excluded by cluster monotonicity, and in addition it excludes $(D(j) \in S_1, D(k) \in S_1)$, both with $D(j) \neq D(k)$, $j\in S_1$, and $k \in S_2$. The classical monotonicity assumption with a single binary treatment and instrument has similar implications: It excludes the response type ($D(j)=k,D(k)=j$), which represents all individuals for whom $Z = j$ makes it less likely to take up treatment $j$.

The plausibility of Assumption~\ref{ass:mono} depends on how treatments are grouped into clusters and how individuals respond to particular encouragements. Suppose that $S_1=\{1,2\}$ is a STEM cluster with Physical sciences (1) and Engineering (2), and $S_2=\{3,4\}$ a non-STEM cluster. Assumption~\ref{ass:mono} allows a STEM offer to induce enrollment in any STEM field, provided that the non-STEM offer induces enrollment in a non-STEM field. For example, $(D(1)=2,D(3)=4)$ is allowed: a student may enroll in Engineering when offered Physical sciences if they enroll in a non-STEM field following a non-STEM offer. 

The assumption instead rules out a STEM offer inducing enrollment in a different STEM field from the one chosen following a non-STEM offer. For example, it rules out $(D(1)=1,D(3)=2)$, where a student enrolls in Physical sciences when offered Physical sciences but in Engineering when offered a non-STEM field. If treatments within a cluster are similar and switching costs are lower within than across clusters, such response types may be less likely: a student who enrolls in Physical sciences following a Physical sciences offer may also do so when entering STEM is more costly, yielding $(D(1)=1,D(3)=1)$, while a student able to reach Engineering following a non-STEM offer may also do so following a Physical sciences offer, yielding $(D(1)=2,D(3)=2)$.

\Cref{tab:shifts2} shows that after invoking Assumption~\ref{ass:mono}, the negative terms and $B_1$ are removed from the decomposition in \Cref{lem:decomp}. However, the types $(D(j)\in S_2, D(k)\in S_2)$ with $D(j) \neq D(k)$ and the corresponding term $B_2$ remain. The following assumption also excludes these types.
\begin{assumption}[Irrelevance]\label{ass:irrelevance}
If $D_l(j)=D_l(k)=0$ for all $l\in S_a$, then $D_n(j)=D_n(k)$ for all $n \notin S_a$, $j \in S_a$, and $k \in S_b$. With two clusters, $a=1$ and $b=2$.
\end{assumption}
This assumption rules out all individuals for whom $D(j)\neq D(k)$ and both $D(j)$ and $D(k)$ are outside $S_1$. This excludes the types $(D(j) \in S_2, D(k) \in S_2)$ with $D(j) \neq D(k)$, $j\in S_1$, and $k \in S_2$. Since \Cref{ass:irrelevance} serves to identify a weighted average effect across treatment clusters, it is generally less restrictive than irrelevance assumptions designed to isolate effects between unclustered treatments, as in \citet{kirkeboen2016field}. For example, with $S_1=\{1,2\}$, $S_2=\{3,4\}$, $j=1$, and $k=3$, Assumption~\ref{ass:irrelevance} allows $(D(1)=2,D(3)=4)$. An irrelevance restriction designed to isolate the unclustered effect of treatment 1 relative to treatment 3 rules out such a response.

\Cref{tab:shifts2} shows that, under Assumptions~\ref{ass:mono} and \ref{ass:irrelevance}, the only response types with $D(j) \neq D(k)$ are those satisfying $(D(j)\in S_1, D(k)\in S_2)$ with $j\in S_1$ and $k \in S_2$. This allows the instrument $Z=j$ ($Z=k$) with initial encouragement for treatment $j$ ($k$) to also induce take up of treatment $l$ ($m$), provided that both $j$ and $l$ ($k$ and $m$) belong to the same cluster. Under these assumptions, $\beta$ identifies a weighted average of LATEs:

\begin{theorem}[Causal interpretation IV estimand]\label{thm:clate}\ \\
Under Assumptions~\ref{ass:standardIV}-\ref{ass:irrelevance}, it holds for ${\beta}$ in \eqref{eq:beta2} that
\begin{align}
    &{\beta} = 
     \sum_{j \in S_1} \sum_{k \in S_2} \sum_{l \in S_1}   \sum_{m \in S_2}
     \lambda_{jklm}\Delta_{jklm},
\end{align}
with the LATEs $\Delta_{jklm}$ and weights $\lambda_{jklm}$ defined as
\begin{align}
    \Delta_{jklm} =& \mathbb{E}[Y(l)-Y(m)|D(j)=l,D(k)=m], \label{eq:late}\\
    \lambda_{jklm} =& \frac{\omega_{jk}\mathbb{P}\left[D(j)=l,D(k)=m\right]
    }{\sum\limits_{j\in S_a}  \sum\limits_{k\in S_b}\omega_{jk}\sum\limits_{l\in S_a}  \sum\limits_{m\in S_b}
    \mathbb{P}\left[D(j)=l,D(k)=m\right]
    },\label{eq:weight}\\
    \omega_{jk}=&\mathbb{P}[Z=j|Z\in S_a]\mathbb{P}[Z=k|Z\in S_b], \label{eq:probz}
\end{align}
where $a=1$ and $b=2$.
\end{theorem}
Since $\sum_{j \in S_1} \sum_{k \in S_2}  \sum_{l \in S_1} \sum_{m \in S_2}\lambda_{jklm}=1$ and $0\leq\lambda_{jklm}\leq1$, ${\beta}$ is a weighted average of LATEs $\Delta_{jklm}$. The weighted average is taken over all possible pairs of treatments with one treatment from $S_1$ and one from $S_2$, and for each of those treatment pairs all possible pairs of instruments between $S_1$ and $S_2$. 

The LATE $\Delta_{jklm}$ equals the causal effect of treatment $l \in S_1$ compared to treatment $m \in S_2$, for the response types $(D(j)=l,D(k)=m)$. Although \Cref{thm:clate} shows that under Assumptions~\ref{ass:standardIV} to \ref{ass:irrelevance} the weighted average of the LATEs is identified, the LATEs $\Delta_{jklm}$ are not necessarily identified under these assumptions. 
% discuss weight
The weights $\lambda_{jklm}$ show which LATEs $\Delta_{jklm}$ contribute the most to ${\beta}$. The weight $\lambda_{jklm}$ represents the normalized proportion of response types $(D(j)=l,D(k)=m)$ with $j,l \in S_1$ and $k,m \in S_2$. The weight is normalized by the cumulative proportion of all these response types.

% R2 comment 1
Theorem~\ref{thm:clate} delivers a transparent expression for the IV estimand in terms of LATEs using that treatment and instrument have the same number of values and the same labels. This restriction can be weakened if instrument values can be partitioned into sets $T_c$ associated with treatment clusters $S_c$, without requiring $|T_c|=|S_c|$ or the same labels. The results then extend by replacing instrument indices $j\in S_c$ by $z\in T_c$. If there are multiple instruments at the treatment cluster level with two clusters, the IV specification is overidentified. This changes the weighting structure in Theorem~\ref{thm:clate}, and a causal interpretation generally requires additional restrictions of the type studied by \citet{bhuller20242sls}. 

\subsection{The TSLS estimand with multiple treatment clusters}

\sloppy Suppose we have $c=1,\dots,C$ mutually exclusive clusters of treatments $S_c$. Define the set $S=\{ S_c\}_{c=1}^C$ that includes all clusters. Define the $C$-dimensional vectors $\mathbf{\tilde{Z}_b}=(1,\tilde{Z}_1,\dots\tilde{Z}_{b-1},\tilde{Z}_{b+1},\dots,\tilde{Z}_{C})'$ and $\mathbf{\tilde{D}_b}=(1,\tilde{D}_1,\dots,\tilde{D}_{b-1},\tilde{D}_{b+1},\dots,\tilde{D}_{C})'$, where $\tilde{Z}_c=\sum_{n \in S_c} Z_n$ and $\tilde{D}_c=\sum_{n \in S_c} D_n$. Consider the TSLS estimand
\begin{align}\label{eq:betaC}
    \mathbf{\beta_b} = \mathbb{E}[\mathbf{\tilde{Z}_b}\mathbf{\tilde{D}_b}']^{-1}\mathbb{E}[\mathbf{\tilde{Z}_b}Y],
\end{align}
where $\mathbf{\beta_b}=(\beta_{0b},\beta_{1b},\dots,\beta_{b-1,b},\beta_{b+1,b}\dots,\beta_{Cb})$. Note that $\beta_{12}$ equals $\beta$ in \eqref{eq:beta2} if $C=2$. 
With more than two treatment clusters, the relevance condition in \Cref{ass:relevance} has to be generalized to the condition that $\mathbb{E}[\mathbf{\tilde{D}_b}\mathbf{\tilde{Z}_b}']$ has full rank. 

We assume \eqref{eq:betaC} is considered if the effect of taking treatment from cluster $S_c$ compared to taking treatment from cluster $S_b$ is of interest, for $c=1,\dots,C$. For instance, the clusters can represent fields of study aggregated to broader study areas. Note that $\mathbf{\beta_b}$ can be considered for any $b=1,\dots,C$, and hence comparing to treatment cluster $S_b$ is without loss of generality.

% Explain response types.
\Cref{tab:shiftsC} describes all response types with $D(j) \neq D(k)$ and more than two treatment clusters. The first four types are similar to the ones in \Cref{tab:shifts2} with $a=1$ and $b=2$: These individuals only take treatment from $S_a$ and/or $S_b$ if the instrument is fixed at values from the clusters $S_a$ and $S_b$. With more than two treatment clusters, we can define five additional response types to instrument values $j\in S_a$ and $k\in S_b$. First, $(D(j)\notin \{S_a, S_b\}, D(k)\in S_a)$ are individuals for whom $Z \in S_a$ makes it less likely to take treatment from $S_a$, and are thus ruled out by \Cref{ass:mono}. Second, the types 
      $(D(j)\in S_b, D(k)\notin \{S_a, S_b\})$,
      $(D(j)\notin \{S_a, S_b\}, D(k)\in S_b)$, and
      $(D(j)\notin \{S_a, S_b\}, D(k)\notin \{S_a, S_b\})$
represent individuals for whom $D(j)\neq D(k)$ and both $D(j)$ and $D(k)$ are outside $S_a$. These are ruled out by \Cref{ass:irrelevance}. Third, $(D(j)\in S_a, D(k) \notin \{S_a, S_b\})$ are types who are not prevented by Assumptions~\ref{ass:mono} and \ref{ass:irrelevance}, and therefore pose an additional identification problem in the multiple treatment cluster setting.

\begin{table}[t!]
    \centering
    \caption{Response types with multiple treatment clusters}
    \label{tab:shiftsC}
    \renewcommand{\arraystretch}{0.85}  
    \begin{tabular}{ll}
    \toprule\toprule
      Response types with $D(j)\neq D(k)$   & Excluded by \\
      \midrule
      $(D(j)\in S_a, D(k)\in S_b)$   & \\
      $(D(j)\in S_b, D(k)\in S_a)$   & \Cref{ass:mono} \\
      $(D(j)\in S_a, D(k)\in S_a)$   & \Cref{ass:mono}\\
      $(D(j)\in S_b, D(k)\in S_b)$   & \Cref{ass:irrelevance}\\ 
      \midrule
      $(D(j)\notin \{S_a, S_b\}, D(k)\in S_a)$   & \Cref{ass:mono} \\
      $(D(j)\in S_b, D(k)\notin \{S_a, S_b\})$   & \Cref{ass:irrelevance}\\
      $(D(j)\notin \{S_a, S_b\}, D(k)\in S_b)$   & \Cref{ass:irrelevance} \\
      $(D(j)\notin \{S_a, S_b\}, D(k)\notin \{S_a, S_b\})$   & \Cref{ass:irrelevance}\\
      $(D(j)\in S_a, D(k) \notin \{S_a, S_b\})$   & \Cref{ass:control} \\ 
      \bottomrule\bottomrule
    \end{tabular}
    \begin{tablenotes}
    \footnotesize
    \item[] Notes: For $a\neq b$, $j\in S_a$, and $k\in S_b$, the rows are mutually exclusive and exhaustive among response types satisfying $D(j)\neq D(k)$. For a given comparison cluster $S_b$, the table applies separately to each $S_a$, $a\neq b$, and Assumptions~\ref{ass:mono}--\ref{ass:control} are imposed for each such pair $(S_a,S_b)$.
    \end{tablenotes}
\end{table}

To prevent these response types, we introduce the following additional assumption.
\begin{assumption}[Control group]\label{ass:control}
If there is an $l\in S_a$ for which $D_l(j)=1$, then $D_n(k)=0$ for all $j \in S_a$, $k \in S_b$, and $n \notin \{S_a, S_b\}$.
\end{assumption}
This assumption states that obtaining a treatment in $S_a$ when the instrument is in $S_a$ implies that no treatment outside $S_a$ or $S_b$ can be obtained when the instrument is in $S_b$. It follows that the types $(D(j)\in S_a, D(k) \notin \{S_a, S_b\})$ with $j \in S_a$ and $k \in S_b$ are excluded.

\Cref{ass:control} may hold in settings with one or more control groups among the $J$ treatments. Without loss of generality, we assume these control groups to be in $S_b$. Individuals may obtain the control when not encouraged to enroll in any other treatment than the control. For instance, \cite{duflo2007using} discuss randomized control trials with multiple treatment arms in development economics, where the control group typically receives no intervention. \cite{behaghel2013robustness} analyze a job market activation program with multiple treatment arms where the control group receives standard services. In the absence of a control group, the literature follows \citet{kirkeboen2016field} and \citet{dahl2023high}: They assume that individuals have a preference ranking over treatments, which is revealed before $Z$ is determined. Let $r_m$ denote the treatment that received rank $m$. Then \Cref{ass:control} may hold in the subsample of individuals whose next-best treatment, $r_2$, belongs to $S_b$. Note that \Cref{ass:control} allows for $D(j)=D(k)$ and does not require $D(k)=k$ for any response types, which makes it less restrictive than the next-best assumption formulated at the unclustered treatment level by \cite{kirkeboen2016field}. \cite{bhuller20242sls} also discuss that the latter assumption can be relaxed to allow for $D(j)=D(k)$. 

Under Assumptions \ref{ass:mono} to \ref{ass:control}, the only response types with $D(j) \neq D(k)$ are those satisfying $(D(j)\in S_a, D(k)\in S_b)$ with $j\in S_a$ and $k \in S_b$. As in the two cluster case, this allows the instrument $Z=j$ ($Z=k$) with initial encouragement for treatment $j$ ($k$) to also induce take up of other treatments within the same cluster. Hence, $\beta_{ab}$ in \eqref{eq:betaC} has a similar causal interpretation as $\beta$ in \eqref{eq:beta2}, and boils down to the same expression with $C=2$:

\begin{theorem}[Causal interpretation TSLS estimand]\label{thm:clateC}\ \\
Under Assumptions~\ref{ass:standardIV}-\ref{ass:control}, it holds for $\mathbf{\beta_b}=(\beta_{0b},\beta_{1b},\dots,\beta_{b-1,b},\beta_{b+1,b},\dots,\beta_{Cb})'$ in \eqref{eq:betaC} that
\begin{align}
    &\beta_{ab} = 
     \sum_{j \in S_a} \sum_{k \in S_b} \sum_{l \in S_a}   \sum_{m \in S_b}
     \lambda_{jklm}\Delta_{jklm},
\end{align}
with $a=1,\dots,b-1,b+1,\dots,C$, and $\lambda_{jklm}$ and $\Delta_{jklm}$ defined in \eqref{eq:late}-\eqref{eq:probz}.
\end{theorem}
The proof is deferred to \Cref{A:proof_clate}. From hereon, we present results for an arbitrary number of clusters $C$, and note that the results hold for $C=2$ without \Cref{ass:control}.

% why we provide new and important results
Theorems~\ref{thm:clate} and \ref{thm:clateC} have direct implications for IV estimation with treatment clusters. Applying IV at the cluster level without specifying assumptions at the unclustered treatment level leaves violations unnoticed \citep{imbens2015causal}, and complicates the discussion of the validity of these assumptions. Our results allow researchers to precisely consider the validity of the required assumptions. Furthermore, they show that the parameter targeted by IV estimation at the cluster level has a well-defined causal interpretation as a weighted average of LATEs under appropriate assumptions, clarifying which LATEs contribute and the weight they receive. In Appendix~\ref{A:Weights}, we provide three strategies for obtaining further insights into the components of this weighted average.

\section{Testable conditions}\label{sec:individual}

The estimand in \eqref{eq:betaC} only requires data at the treatment cluster level. If data at the unclustered treatment level is available, necessary conditions of the identifying assumptions can be tested. It follows from \Cref{thm:clateC} that in the absence of certain response types, the estimand in \eqref{eq:betaC} has a causal interpretation. \Cref{tab:shiftsC} summarizes which types are excluded by the underlying assumptions: Only types described by $D(j) \in S_a$ with $j \in S_a$ and $D(k) \in S_b$ with $k \in S_b$ are allowed. This results in the following necessary conditions.

\begin{proposition}[Necessary conditions causal interpretation]\label{prop:nec_cond}\ \\
Let $A$ be any subset of the support of $Y$, under \Cref{ass:standardIV} it holds that
    \begin{align}\label{eq:phi}
    \phi_{ljk}^A&=\mathbb{E}[\mathbb{I}_{Y\in A} D_l|Z=j]-\mathbb{E}[\mathbb{I}_{Y\in A} D_l|Z=k]\\ \notag
    &=\mathbb{P}[Y(l)\in A,D(j)=l,D(k)\neq l]-\mathbb{P}[Y(l)\in A,D(j)\neq l,D(k)=l],
\end{align}
with the necessary conditions for the causal interpretation of $\beta_{ab}$ in \Cref{thm:clateC}:
\begin{enumerate}
    \item  $\phi_{ljk}^A\geq 0$ if $l,j \in S_a$, $k \in S_b$, and \Cref{ass:mono} holds.
    \item  $\phi_{ljk}^A\leq 0$ if $l,k \in S_b$, $j\in S_a$, and Assumptions~\ref{ass:mono} and \ref{ass:irrelevance} hold. 
    \item  $\phi_{ljk}^A= 0$ if $l\notin \{S_a,S_b\}$, $j \in S_a$, $k \in S_b$, and Assumptions~\ref{ass:mono}-\ref{ass:control} hold. 
\end{enumerate}
\end{proposition}

%R2C1:
Suppose that $S_1=\{1,2\}$, $S_2=\{3,4\}$, $j=1$, and $k=3$, and that Assumption~\ref{ass:standardIV} holds. \Cref{prop:nec_cond} requires $\phi^A_{1,1,3}\ge0$ and $\phi^A_{2,1,3}\ge0$ under Assumption~\ref{ass:mono}, while the corresponding quantities for treatments 3 and 4 must be nonpositive under Assumptions~\ref{ass:mono} and \ref{ass:irrelevance}. Hence, the unclustered data may reveal violations that would be invisible from the cluster level data alone. 

Depending on the support of $Y$, the conditions in \Cref{prop:nec_cond} may pose an infinite number of (in)equalities. \cite{mourifie2017testing} show that such conditions can be written as conditional moment inequalities with $Y$ as the conditioning variable instead of $Z$, allowing them to be tested using the intersection bounds framework of \cite{chernozhukov2013intersection}. \cite{bai2024sharp} follow the same strategy for testing IV assumptions with multiple unordered treatments without clusters.

The Stata package \textit{clrtest} by \cite{ chernozhukov2015implementing} enables joint testing of the conditional moment inequalities within each estimation sample. \cite{mourifie2017testing} discuss the implementation of the multiple testing procedure of \citet{holm1979simple} when the \textit{clrtest} is used to test the conditions across a large number of samples. Statistical power may be further improved by testing the conditions across covariate-specific subsamples using machine learning \citep{farbmacher2022instrument}.

Failure to reject the necessary conditions in \Cref{prop:nec_cond} is not sufficient to conclude that Assumptions~\ref{ass:standardIV} and \ref{ass:mono} to \ref{ass:control} hold. For instance, under \Cref{ass:standardIV}, it may be the case that $\mathbb{P}[Y(l)\in A,D(j)=l,D(k)\neq l]>\mathbb{P}[Y(l)\in A,D(j)\neq l,D(k)=l]>0$, which violates \Cref{ass:mono} while $\phi_{ljk}^A\geq 0$. By contrast, a rejection of the necessary conditions implies that the assumptions required for the causal interpretation of $\beta_{ab}$ are violated. However, the decomposition of the IV estimand in \Cref{lem:decomp} implies that the resulting bias may be small. For instance, the response types that violate the assumptions may receive little weight in the IV estimand, or their corresponding LATEs are similar to those included in the causal estimand.

Researchers may use rejections of specific necessary conditions to find clusters that do meet the assumptions, for instance by sequentially testing the necessary conditions. However, this approach introduces a pre-testing problem when inferring the treatment effect estimates, and may yield classifications for which the assumptions cannot be rejected, rather than classifications for which the assumptions actually hold. Researchers may therefore prefer to use the falsification tests to check the validity of a classification motivated by the empirical context or economic theory.

\section{Fuzzy RDD with treatment clusters}\label{sec:covariates}

This section provides an expression for the TSLS estimand with treatment clusters in a fuzzy regression discontinuity design. Fuzzy RDDs are commonly used in settings with multiple unordered treatments and revealed preference rankings over treatments. This is also the setting in Section~\ref{sec:Application}. 

In a fuzzy RDD, the instrument $Z$ is determined by the value of a continuous running variable lying on either side of a fixed cutoff. We analyze the setting where the running variable induces encouragement toward the multiple unordered treatments as follows:
\begin{align}\label{eq:binaryZ}
    Z = \left\{
    \begin{array}{ll}
        r_1 & S \geq 0, \\
        r_2 & S < 0,
    \end{array}
\right.
\end{align}
where $r_m$ is the treatment receiving rank $m$, and $S$ is the running variable minus the cutoff, with a density that is continuous and strictly positive at $S=0$. 

The fuzzy RDD replaces Assumption~\ref{ass:standardIV}(a) by the local continuity assumption \citep*{dong2018alternative,hahn2001identification}:
\begin{assumption}[Continuity]\label{ass:RD}
$Y(d,z)$ and $D(z)$ are continuous in $S$ at $S=0$ for all $d,z$.
\end{assumption}
This assumption implies that individuals just below and above the cutoff have similar potential outcomes. We control for smooth variation in the running variable with a polynomial function of the running variable $f_h(S)$ with $f_h(0)=0$ within a bandwidth $h$. This ensures that the fuzzy RDD estimand is identified from the discontinuity at the cutoff, rather than also capturing smooth relationships between the running variable and $D$ or $Y$ \citep{imbens2008regression}.

We extend the TSLS estimand in \eqref{eq:betaC} to the RDD. Define the $p$-dimensional vector $\mathbf{X}_h=(\mathbf{F}', \mathbf{B}_h')$, where $\mathbf{F}$ includes the fixed effects $f_{jk}$ that equal one if $(r_1,r_2)=(j,k)$ and zero otherwise, with $j \neq k \in S_b$, and $\mathbf{B}_h'$ includes the basis functions of the polynomial function $f_h(S)$. The TSLS estimand equals
\begin{align}\label{eq:betaCx}
(\mathbf{\delta_b^h}', \mathbf{\beta_b^h}')'
    = \mathbb{E}_h[\mathbf{\tilde{Z}_{b,h}}\mathbf{\tilde{D}_{b,h}}{}']^{-1}\mathbb{E}_h[\mathbf{\tilde{Z}_{b,h}}Y],
\end{align}
where $\mathbf{\tilde{Z}_{b,h}}=(\mathbf{X}_h',\mathbf{\tilde{Z}_b})'$, $\mathbf{\tilde{D}_{b,h}}=(\mathbf{X}_h',\mathbf{\tilde{D}_b})'$, and $\mathbb{E}_h[A]=\mathbb{E}[A\mid |S|\leq h, r_2 \in S_b]$. If each cluster contains only one treatment, this boils down to the fuzzy RDD estimand with unclustered treatments, and hence the results in this section also hold without treatment clusters.  
The estimand in \eqref{eq:betaCx} corresponds to the TSLS estimand of the equations:
\begin{align}\label{eq:2stage_rdd}
    Y &= \sum_{k \in S_b}\sum_{l \neq k} \nu_{lk}^hf_{lk}+\sum_{a \neq b}\beta_{ab}^h\tilde{D}_a+ f_{h,b} (S)  + \varepsilon_{h,b},\\ \label{eq:1stage_rdd}
    \tilde{D}_a &= \sum_{k \in S_b}\sum_{l \neq k} \eta_{alk}^hf_{lk}+\sum_{c\neq b} \pi_{acb}^h \tilde{Z}_c +f_{h,ab}(S)+ u_{h,ab}, \quad \text{\textit{for all }} a \neq b.
\end{align}
Additional controls can be taken into account with saturated specifications that control for covariates nonparametrically \citep{blandhol2022tsls}, or by double/debiased machine learning methods \citep{chernozhukov2018double}.

Theorem \ref{thm:clate_rdd} shows that the TSLS estimand in the fuzzy RDD has a causal interpretation under the local continuity assumption.
\begin{theorem}[Causal interpretation fuzzy RDD estimand]\label{thm:clate_rdd}\ \\
Under Assumptions~\ref{ass:standardIV}(b) and \ref{ass:relevance}-\ref{ass:RD}, and $f_h(S)$ varying flexibly with the fixed effects, it holds for $\mathbf{\beta_{b}^h}=(\beta_{0b}^h,\beta_{1b}^h,\dots,\beta_{b-1,b}^h,\beta_{b+1,b}^h,\dots,\beta_{Cb}^h)'$ in \eqref{eq:betaCx} that
\begin{align}
   \lim_{h \downarrow 0} {\beta_{ab}^h}=\sum_{j \in S_a}\sum_{k \in S_b}\sum_{l \in S_a}\sum_{m \in S_b} \lambda_{jklm} \Delta_{jklm},
\end{align}
with $a=1,\dots,b-1,b+1,\dots,C$ and
the LATEs $\Delta_{jklm}$ and weights $\lambda_{jklm}$ defined as
\begin{align*}
    \Delta_{jklm} =& \mathbb{E}[Y(l)-Y(m)|D(j)=l,D(k)=m,f_{jk}=1,S=0], %\label{eq:late_rdd}
    \\
    \lambda_{jklm} =& \frac{\omega_{jk}\mathbb{P}\left[D(j)=l,D(k)=m|f_{jk}=1,S=0\right]
    }{\sum\limits_{j\in S_a}  \sum\limits_{k\in S_b}\omega_{jk}\sum\limits_{l\in S_a}  \sum\limits_{m\in S_b}
    \mathbb{P}\left[D(j)=l,D(k)=m|f_{jk}=1,S=0\right]
    },%\label{eq:weight_rdd}
    \\
    \omega_{jk}=&\lim_{h \downarrow 0} \mathbb{P}[f_{jk}=1|r_2 \in S_b, |S|\leq h]. %\label{eq:probf}
\end{align*}
\end{theorem}
The proof is deferred to Appendix~\ref{A:clate_rdd}.
% compare with Theorem 2
The fuzzy RDD estimand identifies a weighted average of LATEs, with weights $0\leq\lambda_{jklm}\leq1$ and $\sum_{j \in S_a} \sum_{k \in S_b} \sum_{l \in S_a} \sum_{m \in S_b}\lambda_{jklm}=1$. Similar to \Cref{thm:clateC}, the weights represent a normalized proportion of response types corresponding to each LATE: the individuals near the cutoff with $r_1=j$ and $r_2=k$ and $(D(j) \in S_a, D(k) \in S_b)$ with $j\in S_a$ and $k\in S_b$.

\section{Application with fields of study}\label{sec:Application}

This section illustrates the empirical relevance of our methods in a setting with fields of study as multiple unordered treatments. We consider a natural experiment in the Portuguese public higher education system to study the effect of fields of study on the achieved number of study credit points.  

\subsection{Context and empirical specification}

In the final year of high school, prospective students in Portugal apply to public higher education via a centralized admission process. Only 17 percent of the students use a different enrollment system to enroll into private higher education \citep{pordata}. Applicants apply to a field and institution simultaneously, which we will refer to as a course (\textit{e.g.}, Law at the University of Lisbon). 
Applicants submit a ranking of up to six courses based on their preferences. Each course has a quota of students it can admit and ranks its applicants by a score based on high school and national exam grades. Because courses weight these components differently, application scores are applicant-course specific. Using these rankings and quotas, a deferred acceptance mechanism similar to the college-proposing Gale-Shapley algorithm \citep{gale1962college} matches each applicant to at most one course \citep{dges2019}. Appendix \ref{A:context} describes this mechanism in more detail.

\subsubsection{Course cutoffs, instruments, and next-best fields of study}\label{sec:coursecutoffs}

The admission mechanism generates ex-ante unknown application-score cutoffs for each course. Each cutoff represents the lowest score that qualifies an applicant to receive an offer from that course. We construct the preferred and next-best fields from the course cutoffs, application scores, and submitted rankings. First, we retain applicants who rank courses from at least two different fields. Second, we construct preference blocks by assigning the same rank to consecutive courses in the submitted ranking that belong to the same field. Third, we consider the applicant's matched block and the blocks immediately above and below it. We retain the block below only if the applicant scores above the cutoff for at least one course in that block, ensuring that it represents a feasible alternative. We drop applicants with no available adjacent block. If both adjacent blocks are available, we retain the block above and the matched block.

For these applicants, the higher-ranked retained block defines the preferred field $r_1$, while the lower-ranked block defines the next-best field $r_2$. These fields are local alternatives around the relevant course cutoff and need not correspond to the first two courses or fields in the submitted ranking.

Let $\mathcal{Q}_1$ denote the set of courses in the preferred block, and let $s_q$ and $c_q$ denote the applicant's score and the cutoff for course $q$, respectively. We define the running variable as $S=\max_{q\in\mathcal{Q}_1}\{s_q-c_q\}$. Thus, $S\geq0$ indicates that the applicant clears the cutoff for at least one course in the preferred block, whereas $S<0$ indicates that none of these cutoffs is cleared. The applicant receives an offer from field $j$, and the instrument $Z_j$ equals one, if $j$ is the preferred field and the applicant clears the cutoff for at least one course in that block. The cutoff therefore represents either a single course-specific cutoff or the boundary formed by multiple course-specific cutoffs within the preferred block, and the binary representation in \eqref{eq:binaryZ} applies.

\Cref{tbl:AllocationMechanism} illustrates the construction of the preferred and next-best fields and the running variable. Panel A shows an applicant who ranks two courses in field $j$, followed by two courses in field $k$. The courses in each field form a preference block. The applicant does not clear the cutoff for the first-ranked course in field $j$ but exceeds the cutoff for the second-ranked course by 0.1 points. Because the running variable is the maximum re-centered score among all courses in the preferred block, it equals $S=0.1$. The applicant also clears the cutoff for the highest-ranked course in field $k$, making the remaining scores in that block irrelevant. Field $j$ is therefore the preferred field, field $k$ is the next-best field, and the applicant receives an offer from field $j$.

\begin{table}[t!]
\small
  \begin{center}
    \caption{Illustration of the preferred and next-best fields and the fuzzy RDD}
    \renewcommand{\arraystretch}{0.75}  
    \label{tbl:AllocationMechanism}
    \begin{tabular}{c c c c c c c c c c c }
    \toprule\toprule
       Ranking  & &  Field & Institution  &  & Score & Cutoff & & Block  & $Z$ & $S$\\
 		\midrule
 	    \multicolumn{11}{c}{Panel A: Course ranking} \\
    [.5em]
 	  1st & & $j$ & B & & $<$ 190 & 190 & & 1 & $j$ & 0.1 \\ 
      2nd & & $j$ & A & &  180.1  & 180& &  1 & $j$ & 0.1\\
      3rd & & $k$ & B & & $\geq$ 170 & 170 & & 2 & $j$ & 0.1\\
      4th & & $k$ & A  & & $\cdot$&$\cdot$ & & 2 & $j$ & 0.1\\
      5th & & $l$ & E & & $\cdot$&$\cdot$ & & 3 & $j$ & 0.1 \\
      6th & & $m$ & E & & $\cdot$& $\cdot$& & 4 & $j$ & 0.1\\
		[.5em]
        \midrule
        \multicolumn{11}{c}{Panel B: The fuzzy RDD} \\
    [.5em]
       Preferred ($r_1$) & & $j$ & A & & 180.1  & 180 & &  1 & $j$ & 0.1\\
      Next-best ($r_2$) & & $k$ & B & & $\geq$ 170 & 170 & &  2 & $j$ & 0.1\\
  	 & & & & & & &  & & & \\
       Preferred ($r_1$) & & $j$ & C & & 179.9  & 180 & &  $\cdot$ & $k$ & -0.1 \\
      Next-best ($r_2$) & & $k$ & D & & $\geq$ 160 & 160& &  $\cdot$ & $k$ & -0.1  \\
       \bottomrule \bottomrule
    \end{tabular}
        \begin{tablenotes}
	\footnotesize
	\item[] Notes: Panel A provides an illustrative example of an applicant's course ranking and the corresponding outcome of the admission mechanism. Panel B illustrates how this ranking and outcome can be used to construct the preferred and next-best fields for a fuzzy RDD. %See the text for details on the construction of the columns Block, $Z$, and $S$.   
\end{tablenotes}
  \end{center}
\end{table}

Panel B summarizes and compares this applicant to another applicant with the same preferred and next-best fields who falls 0.1 points below the relevant cutoff for field $j$ but clears a course cutoff in field $k$. The two applicants therefore have $S=0.1$ and $S=-0.1$ and receive offers from fields $j$ and $k$, respectively. The fuzzy RDD compares applicants immediately above and below these re-centered cutoffs. Because courses combine a field and an institution, crossing a cutoff may change both. Our field effects therefore include effects associated with any accompanying change in institution.

Our preferred and next-best fields may differ from students’ true preferences for three reasons. First, they represent local alternatives around a particular cutoff and need not be the two highest-ranked fields in the submitted ranking. Second, applicants can rank at most six courses, which may induce them to include safe options near the end of their rankings. These first two concerns appear limited in our setting because, for 88\% of the sample, $r_1$ and $r_2$ coincide with the first- and second-ranked fields in the submitted ranking. Third, the college-proposing Gale-Shapley mechanism is not strategy proof from the applicants’ perspective. Applicants may have incentives to misreport their preferences beyond the first-ranked course \citep{roth1982economics}, potentially introducing error into $r_2$. 

\citet{bertanha2024causal} discuss how strategic misreporting of preferences complicates identification of causal parameters that depend on true preferences. They also show that strategic reporting may violate the RD continuity assumption. They develop a partial identification approach robust to such behavior in a sharp RDD but leave identification in a fuzzy RDD with ``multiple compliance types'' to future work. Although we cannot rule out strategic reporting, the balancing and bunching tests reported below do not indicate violations of the RD continuity assumption.

\subsubsection{Data and classifications of fields of study}

We have access to an administrative data set that includes four cohorts in the Portuguese public higher education system. From the application data we observe each applicant's submitted course ranking, application scores, and matched course, on top of all course cutoffs. From the student outcome data we observe a student's enrolled course, whether a student is active in the first, second, or third year of that course, and the achieved number of study credits in each year for that course. Appendix~\ref{A:data} provides further details on the data and sample construction and presents some descriptive statistics.

Study credits are measured using the European Credit Transfer and Accumulation System (ECTS), which facilitates the accumulation and transfer of learning achievements across European universities. ECTS credits represent learning outcomes and the estimated workload required to achieve them. One ECTS credit typically represents 25 to 30 hours of intended workload, while 60 ECTS correspond to one academic year, or 1,500 to 1,800 hours of workload. Four and eight ECTS typically correspond to a medium-sized and large course, respectively. These standards apply across institutions and fields, making ECTS comparable in terms of intended workload, but not in terms of content, difficulty, or quality. 

We interpret achieved ECTS as a measure of study progress rather than a measure of skills or welfare: achieving more credits represents the completion of more intended workload and greater progress toward degree completion, but does not necessarily imply more skill development, enjoyment, or future earnings. Universities in several European countries eventually dismiss students who obtain insufficient credits \citep{sneyers2018interventions}. In those settings, differences in academic progress could affect continued enrollment. However,  dropout in Portugal is low, so this is not the focus of this application. 

The Portuguese education authorities organize courses into 22 fields of study following the 2-digit International Standard Classification of Education (ISCED) from \cite{unesco2015international}. Column (1) of \Cref{tbl:classification} shows the  22 fields of study, which we define as our unclustered treatments. Column (2) classifies the 22 fields into two clusters according to the STEM versus non-STEM distinction from \cite{OECDglance2022}. Column (3) constructs eight clusters using the ISCED 1-digit classification: fields that share the same first digit in their 2-digit classification are grouped together in a broader field. 

\begin{table}
\footnotesize
\begin{adjustwidth}{-1.45cm}{-1.45cm}
    \caption{STEM and ISCED classification of the fields of study}
    \label{tbl:classification}
    \renewcommand{\arraystretch}{0.85}
    \begin{tabular}{lllcccc}
    \toprule\toprule
   Fields of study &  STEM  &ISCED & Sample &\multicolumn{3}{c}{Reject necessary conditions} \\ [-.5em]
     &  &  & size & Fields & STEM &  ISCED  \\
   \midrule 
	Teacher training and education science  &		Non-STEM	&		Education		
    & 1370 & &	&	\\
	Arts	&		Non-STEM  &		Humanities and arts	
    & 1735 &  &	&	\\
	Humanities	&		Non-STEM	&		Humanities and arts 
    & 2875 & 	&	& 	\\
	Social and behavioural science	&		Non-STEM	&		Soc. sci., bus. and law
    & 7014 & 	& 	& 	\\
	Journalism and information	&		Non-STEM	&		Soc. sci., bus. and law 
    & 1313 & 	& 	& 	\\
	Business and administration	&		Non-STEM	&		Soc. sci., bus. and law 
    & 7432 & 	&	&	\\
	Law	&		Non-STEM   &		Soc. sci., bus. and law 
    & 1379 & 	&	&	\\
	Life sciences	&		STEM	&		Science 
    & 4340 & 	& 	& 	\\
	Physical sciences	&		STEM&		Science 
    & 2672 & 		&	& 	\\
	Mathematics and statistics	&		STEM	&		Science 
    & 926 & 	&	& 	\\
	Computing	&		STEM&		Science	
    & 1755 & 	&	& 	\\
	Engineering and engineering trades	&		STEM	&		Eng., man. and constr. 
    & 5214 &   &	 & 	\\
	Manufacturing and processing	&		STEM	&		Eng., man. and constr. 
    & 1102 & 	&	&	\\
	Architecture and building	&		STEM	&		Eng., man. and constr. 
    & 2063 & 	& 	& 	\\
	Agriculture, forestry and fishery	&		Non-STEM	&		Agriculture 
    & 363 & 	&	&	\\
	Veterinary	&		Non-STEM	&		Agriculture 
    & 716 & 	&	&	\\
	Health	&		Non-STEM	&		Health and welfare 
    & 2855 & Reject	& 	& Reject	\\
	Social services	&		Non-STEM	&		Health and welfare 
    & 1264 & 	&	& 	\\
	Personal services	&		Non-STEM	&		Services 
    & 1813 & Reject	&	& Reject	\\
	Transport services	&		Non-STEM	&		Services 
    & 21 & 	&	&	\\
	Environmental protection	&		STEM  &		Services 
    & 982 & 	&	& 	\\
	Security services	&		Non-STEM	&		Services 
    & 68 & 	&	&	\\
    \bottomrule \bottomrule
    \end{tabular}
    \begin{tablenotes}
	\footnotesize
	\item[] Notes: The first three columns show the 22 fields of study, the crosswalk to STEM versus non-STEM from \cite{OECDglance2022}, and the ISCED clusters from \cite{unesco2015international}. The fourth column reports the number of students in our sample with the field in Column (1) as next-best field $k$.  Reject in the final three columns indicates that the necessary conditions in \eqref{eq:phi}, with $k$ corresponding to the field of study in Column (1), are rejected after a Holm correction at the 5\% significance level for the fields of study, STEM clusters, or ISCED clusters.  
    \end{tablenotes}
\end{adjustwidth}
\end{table}

\subsubsection{Empirical specification}

We estimate the fuzzy RD model as specified in \eqref{eq:2stage_rdd}-\eqref{eq:1stage_rdd} using the sample analogue of \eqref{eq:betaCx}. For treatment cluster $S_b$, the estimation sample contains students whose next-best field belongs to $S_b$. For the unclustered field classification, where each cluster contains one field, Column (4) of \Cref{tbl:classification} reports these estimation samples. For the STEM and ISCED classifications, we pool the samples over all next-best fields within the same cluster.

The outcome variable $Y$ equals the number of study credits collected in the second year of a student's course. The treatment variable $\tilde{D}_a=\sum_{j\in S_a} D_j$ equals one if the student's second-year course enrollment corresponds to a field in cluster $S_a$. The instrument $\tilde{Z}_c=\sum_{l\in S_c} Z_l$ equals one if the student prefers a field in cluster $S_c$ and has an application score above one of its course cutoffs. Clustering therefore pools discontinuities around the re-centered course-specific cutoffs. The fixed effect $f_{jk}$ equals one if the preferred and next-best field of the student are $j$ and $k$, respectively.  

The function $f_h(S)$ is approximated by a first-degree polynomial, such that $f_h(S)=\gamma S$, and we do not apply a bandwidth to our sample. \Cref{A:robustness} shows that our results are robust against the inclusion of higher order polynomials and the use of smaller bandwidths. 
Furthermore, we report on balancing and bunching tests leading to the conclusion that we cannot reject the RD continuity assumption. We use heteroskedasticity robust standard errors and apply the \cite{holm1979simple} correction to account for multiple hypothesis tests within each estimation sample.

Because the treatment variable is defined as second-year course enrollment, our estimation sample is restricted to students with observed second-year enrollment. The local continuity assumption then requires that the course offer does not affect selection into this sample near the cutoff. Non-enrollment and dropout are relatively low in Portugal. Approximately 10\% of applicants do not enroll in public higher education in their first year \citep{dges2018}, while approximately 2.5\% of those who enroll drop out before the second year \citep{figueiredo2026balanco}. The latter is consistent with Portugal’s fixed admission quotas, which give little incentive to induce dropout after admission. 
These rates suggest limited attrition, while balancing and bunching tests reveal no significant irregularities around the cutoff. Together, these findings are reassuring, but do not entirely rule out offer-induced selection near the cutoff.

With our treatment definition, the exclusion restriction also requires that course offers affect second-year credits only through the second-year field of enrollment. Two potential violations merit discussion. First, offers may affect the institution without changing the field, in which case differences in institutional characteristics could directly affect credits. Such effects are likely to be limited as Portuguese public institutions  offering the same fields have similar curricula and academic requirements. Second, offers may change first-year field without changing the second-year field. If these earlier experiences directly affect second-year credits, exclusion is violated. A formal analysis of these multiple treatment stages is beyond the scope of this paper, but \citet{angrist2025one} discuss identification in these settings with a single instrumental variable. 

\subsubsection{Identifying assumptions with treatment clusters}\label{sec:identifyingass}

The set of identifying assumptions required for a causal interpretation of the TSLS estimand with treatment clusters depends on the number of clusters. In addition to the standard IV assumptions, the STEM classification with two treatment clusters requires Assumptions~\ref{ass:mono} and \ref{ass:irrelevance}. We discuss the plausibility of these two assumptions in turn. 

First, consider a student whose next-best field is in the non-STEM cluster, such as Personal services, and whose preferred field is in the STEM cluster, such as Physical sciences. A score above the cutoff implies an offer to enroll in Physical sciences rather than Personal services. \Cref{ass:mono} rules out students for whom a Physical sciences offer causes them to leave the STEM cluster. This seems plausible, as it is unlikely that receiving an offer from a STEM field reduces the probability of enrolling in any STEM field. The assumption also rules out a STEM offer inducing enrollment in a different STEM field from the one chosen following a non-STEM offer, such as a student who would enroll in Engineering following a Personal services offer but in Physical sciences following a Physical sciences offer. Second, \Cref{ass:irrelevance} rules out students who enroll in different non-STEM fields depending on whether they receive a Physical sciences or Personal services offer. If receiving an offer from a STEM field does not make a student enroll into any STEM field, it is indeed unlikely that the offer induces enrollment into any non-STEM field instead. 

%more than two clusters
Since the ISCED classification has more than two treatment clusters, identification requires Assumptions~\ref{ass:mono}, \ref{ass:irrelevance}, and \ref{ass:control}. We discuss these assumptions in the context of the same example. Consider the next-best field of study Personal services in the cluster Services and the preferred field Physical sciences in the cluster Science. Similar to the two treatment cluster case, \Cref{ass:mono} ensures that receiving a Physical sciences offer does not make it less likely that a student enrolls into any Science field, and \Cref{ass:irrelevance} prevents students who enroll in different non-Science fields depending on whether they receive a Physical sciences or Personal services offer. \Cref{ass:control} additionally requires that a student enrolling into any Science field after receiving an offer from Physical sciences, does not enroll into any non-Services field when receiving an offer from Personal services. This is likely if the Services cluster represents a control group, as is the case for the sample of students with a next-best field in the Services cluster. 

%when are assumptions met?
The validity of Assumptions~\ref{ass:mono} to \ref{ass:control} depends on whether clusters effectively group similar fields of study, such that students may update preferences and face low switching costs between fields within, but not across, treatment clusters. \cite{wiswall2015determinants} find evidence that student preferences are similar for majors within broad academic groups, suggesting that students initially enrolling in Physical sciences (Personal services) may later adjust their preferences towards related fields in Science (Services). \cite{altonji2016analysis} discuss that major-switching costs depend on a student's education history. \cite{dgesmudanca} describes that course changes in Portugal require an official request that includes potential accreditation of previous coursework. Since related fields share curricula, accreditation is often easier within clusters.

\subsection{Results}

\subsubsection{Two treatment clusters: STEM and non-STEM}

We find that STEM enrollment negatively impacts the number of achieved study credits in the second year. \Cref{tab:stem} shows that based on our estimates, STEM enrollment decreases ECTS on average by 3.15 credits for students preferring STEM over non-STEM in their course ranking. This decrease represents approximately 88 fewer hours of intended workload under STEM compared with non-STEM enrollment. The estimated effect is smaller for students preferring non-STEM over STEM: non-STEM enrollment increases ECTS on average by 1.33 credits for these students. The magnitude of this effect represents approximately 37 additional hours of intended workload. In case students are aware of the costs of STEM enrollment, their preference for and enrollment into STEM may be explained by other benefits, such as skill development or future monetary payoffs.

\begin{table}[t!]
    \centering
    \caption{Treatment effect and treatment take-up estimates with STEM clusters}
    \label{tab:stem}
    \renewcommand{\arraystretch}{0.75}
    \begin{tabular}{ccccc}
    \toprule \toprule
    & \multicolumn{4}{c}{Treatment cluster $S_b$} \\
    \cmidrule(lr){2-5}
    & non-STEM & STEM & non-STEM & STEM \\
    \midrule
    &\multicolumn{2}{c}{Treatment effect estimates} & \multicolumn{2}{c}{Treatment take-up estimates} \\ [-.5em]
    & \multicolumn{2}{c}{} &\multicolumn{2}{c}{for enrollment $\widetilde{D}_a$} \\
Estimate  & -3.15***	&	1.33*** &	0.88***	&	0.82***	\\ [-.5em]
&  (0.54)	&	(0.44) &	(0.01)	&	(0.01)	\\
Observations & 30218 & 19054 & 30218 & 19054 \\
      \bottomrule \bottomrule
    \end{tabular}
    \begin{tablenotes}
    \footnotesize
    \item[] Notes:
    The first two columns report estimates of $\beta_{ab}^h$ from \eqref{eq:2stage_rdd}, while the final two report estimates of $\pi_{acb}^h$ from \eqref{eq:1stage_rdd}, using the STEM classification. For both sets of estimates, the columns correspond to treatment clusters $S_b$. The final row reports the size of the estimation sample comprising students whose next-best field belongs to the corresponding cluster. The estimates of $\beta_{ab}^h$ are obtained using 2SLS and those of $\pi_{acb}^h$ using OLS, with robust standard errors in parentheses. Stars indicate statistical significance at the ${}^{*}10$, ${}^{**}5$, and ${}^{***}1$\% levels. %No Holm correction is needed because each next-best estimation sample involves only one test.
    \end{tablenotes}
\end{table}

% treatment effect interpretation
The estimates can be interpreted as weighted averages of LATEs between fields of study across the two treatment clusters. The STEM and non-STEM clusters include respectively 8 and 14 fields of study. The weighted averages include the LATEs corresponding to all possible pairs of fields across these two clusters, and for each of those field pairs all possible pairs of instruments between the two clusters. It follows that the treatment effect of STEM versus non-STEM (and vice versa) includes $8\times14\times8\times14=12,544$ LATEs and corresponding weights. 

The large and precisely estimated treatment take-up effects provide strong evidence that Assumption~\ref{ass:relevance} holds for the STEM classification. \Cref{tab:stem} shows that, for students preferring a field in STEM over non-STEM, scoring above the cutoff and thereby receiving an offer from a STEM field increases the probability of enrollment in a STEM field by 88 percentage points. Conversely, for students preferring non-STEM over STEM, receiving a non-STEM offer increases the probability of non-STEM enrollment by 82 percentage points. 

% test necessary conditions
Since we have access to the data at the field of study level, we can also test the necessary conditions in \Cref{prop:nec_cond} for a causal interpretation of the treatment cluster effects. We jointly test the (in)equalities in \Cref{prop:nec_cond} within each sample defined by a preferred and next-best field using the \textit{clrtest} package in Stata. Appendix~\ref{A:testingnecessary} provides further implementation details. Column (5) of \Cref{tbl:classification} summarizes the testing results. After applying a Holm correction, we find no statistically significant violations of the necessary conditions for a causal interpretation of the estimated treatment cluster effects.

This result may be explained by most students having response types that can be described by $(D(j)=j,D(k)=k)$ with $j \in S_a$ and $k \in S_b$. Define $\pi_{ljk}^h$ as the coefficients $\pi_{acb}^h$ in \eqref{eq:1stage_rdd} if each cluster contains only one treatment. The treatment take-up estimates for $\pi_{ljk}^h$ indeed suggest that most students enroll into their next-best or preferred field when they receive an offer from the next-best or preferred field, respectively. In particular, the average estimate of $\pi_{jjk}^h$ is 0.87, whereas the average absolute value of $\pi_{ljk}^h$ for $l\neq j,k$ is 0.01. Together, these estimates leave little scope for response types that violate Assumptions~\ref{ass:mono} to \ref{ass:control} and reduce concerns about violations of the exclusion restriction. The full set of treatment take-up estimates at the field of study level is provided as supplementary data in the Supporting Information.

\subsubsection{Multiple treatment clusters: Eight ISCED clusters}

\Cref{tab:isced_ss} shows the treatment cluster effect estimates across the eight ISCED clusters. After the Holm correction within each estimation sample, four estimates are statistically significant at the 5\% level, compared with twelve before the correction. Their average absolute values are 4.13 and 3.44 ECTS, respectively. These effect sizes imply that some clustered fields can increase or decrease academic progress by the equivalent of approximately 100 hours of intended workload, or a medium-sized course.

\begin{table}[t!]
    \begin{adjustwidth}{-1.45cm}{-1.45cm}
    \caption{Treatment effect estimates with ISCED clusters}
    \label{tab:isced_ss}
     \renewcommand{\arraystretch}{0.75}
   \begin{tabular}{lcccccccccc}
    \toprule \toprule
Enrollment $\widetilde{D}_a$  & \multicolumn{8}{c}{Treatment cluster $S_b$}\\
\cmidrule(lr){1-1}\cmidrule(lr){2-9}
   & Edu- & Hum. & Soc. sci., & Science & Eng., & Agri- & Health & Ser- \\ [-.5em]
& cation & and& bus. and& & man. and & culture& and & vices\\  [-.5em]
&  & arts& law& & constr. & & welfare & \\ 
\midrule    
Education	&		&	-1.15	&	2.43	&		&		&		&	-0.79	&	-1.10	\\ [-.5em]
	&		&	(2.22)	&	(1.75)	&		&		&		&	(1.21)	&	(3.91)	\\
Humanities and arts	&	-4.33*	&		&	-0.16	&	1.71	&	1.34	&		&	-3.81	&	-1.73	\\ [-.5em]
	&	(2.02)	&		&	(0.94)	&	(3.75)	&	(1.76)	&		&	(2.74)	&	(2.71)	\\
Soc. sci., bus. and law	&	-2.66*	&	1.01	&		&	3.49**	&	2.38	&		&	-3.51***	&	-2.36	\\ [-.5em]
	&	(1.08)	&	(0.80)	&		&	(1.13)	&	(1.42)	&		&	(0.80)	&	(1.13)	\\
Science	&		&	-4.52	&	-2.23	&		&	-2.16	&	-0.62	&	-2.32	&	1.35	\\ [-.5em]
	&		&	(6.41)	&	(1.32)	&		&	(0.92)	&	(4.08)	&	(1.15)	&	(2.52)	\\
Eng., man. and constr.	&		&	-5.73**	&	-1.78	&	-0.56	&		&	0.24	&	-3.81**	&	-4.36	\\ [-.5em]
	&		&	(1.86)	&	(1.21)	&	(0.79)	&		&	(3.53)	&	(1.23)	&	(2.68)	\\
Agriculture	&		&		&	-5.36	&	-1.09	&	-4.35	&		&	2.30	&	-3.36	\\ [-.5em]
	&		&		&	(3.09)	&	(1.72)	&	(2.28)	&		&	(1.85)	&	(5.19)	\\
Health and welfare	&	0.31	&	3.24	&	1.71	&	-0.33	&	1.12	&	-2.18	&		&	-1.14	\\ [-.5em]
	&	(1.61)	&	(1.86)	&	(0.85)	&	(0.79)	&	(0.77)	&	(1.90)	&		&	(1.41)	\\
Services	&	-2.56	&	3.59	&	2.69*	&	3.53	&	6.47	&	-13.67	&	-0.64	&		\\ [-.5em]
	&	(2.30)	&	(2.05)	&	(1.05)	&	(2.99)	&	(2.84)	&	(8.15)	&	(2.05)	&		\\
 Observations  & 1370 & 4610 & 17138 & 9693 & 8379 & 1079 & 4119 & 2884\\ 
      \bottomrule \bottomrule
    \end{tabular}
    \begin{tablenotes}
    \footnotesize
    \item[] Notes: This table reports estimates of $\beta_{ab}^h$ from \eqref{eq:2stage_rdd}, using the ISCED classification. The columns correspond to treatment clusters $S_b$ and the rows to treatment clusters $S_a$. The final row reports the size of the estimation sample comprising students whose next-best field belongs to the corresponding cluster $S_b$. The estimates are obtained using 2SLS, with robust standard errors in parentheses. Empty off-diagonal entries indicate pairs of treatment clusters for which all combinations of preferred and next-best fields were excluded because of insufficient observations (see \Cref{A:data} for details). Stars indicate statistical significance at the ${}^{*}10$, ${}^{**}5$, and ${}^{***}1$\% levels after applying a Holm correction for the number of tests within each estimation sample.
    \end{tablenotes}
\end{adjustwidth}
\end{table}

For the ISCED classification, we also find strong evidence that Assumption~\ref{ass:relevance} holds. \Cref{tab:isced_fs} in Appendix~\ref{A:takeup} reports the treatment take-up estimates for enrollment in the cluster containing the preferred field. Consistent with the evidence above that response types are predominantly characterized by $(D(j)=j,D(k)=k)$, these estimates are large and precise, generally ranging from 0.80 to 0.90. 

% test
We test the necessary conditions in \Cref{prop:nec_cond} for the eight clusters and reject a causal interpretation of the treatment cluster effect estimates. Column (6) of \Cref{tbl:classification} shows that we reject the necessary conditions for the fields Health and Personal services. These rejections do not affect the conditions for the STEM classification. For instance, the rejected conditions in the sample of next-best Personal services correspond to the preferred field Humanities, and both Personal services and Humanities belong to the non-STEM cluster. Appendix \ref{A:additionalresults} provides additional empirical results for both the STEM and ISCED classifications, also using the data at the field of study level.

\section{Conclusion}\label{sec:conc}

Many modern policies involve multiple unordered treatments, for which many pairwise treatment comparisons are possible. A common approach is to aggregate these treatments into clusters and apply instrumental variable (IV) methods at the treatment cluster level.

This paper shows that IV methods with treatment clusters identify weighted averages of pairwise effects between treatments across clusters. The required assumptions are formulated at the unclustered treatment level and hence explicit about possible violations. We discuss how practitioners can justify these assumptions in the presence of treatment clusters and show that the assumptions are less strict compared to the assumptions required for identifying the pairwise effects between unclustered treatments. We also derive necessary conditions for these assumptions, and propose falsification tests for settings in which data at the unclustered treatment level is available.

We illustrate the empirical relevance of our methods by analyzing the causal effects of clusters of fields of study using a natural experiment in the Portuguese public higher education system. We consider the STEM versus non-STEM classification and eight clusters defined according to the International Standard Classification of Education. We find meaningful effects of enrollment in different clusters on study progress. Using data at the field of study level, we test the necessary conditions for causal interpretation of the treatment cluster effects. 

\bibliographystyle{chicagoa}
\bibliography{Library.bib}

\clearpage
\appendix

\section{Proofs theoretical results}

\subsection{Proof Lemma~\ref{lem:decomp}}\label{A:proof_decomp}
The denominator of the IV estimand in \eqref{eq:beta2} can be written as
\begin{align} 
    \Delta\mathbb{E}[\tilde{D}_1|\tilde{Z}_1] =&  
       \sum_{n\in S_1} \mathbb{E}[D_n|\tilde{Z}=1]-  \sum_{n\in S_1}\mathbb{E}[D_n|\tilde{Z}=2] \\
      =& \sum_{j\in S_1} \sum_{n\in S_1} \mathbb{E}\left[D_n|Z=j \right]\mathbb{P}[Z=j|Z\in S_1] - 
       \sum_{k\in S_2} \sum_{n\in S_1}\mathbb{E}\left[D_n|Z=k\right]\mathbb{P}[Z=k|Z\in S_2]\notag\\
       =& \sum_{j\in S_1} \sum_{n\in S_1} \mathbb{E}\left[D_n(j)\right]\mathbb{P}[Z=j|Z\in S_1] - 
       \sum_{k\in S_2} \sum_{n\in S_1}\mathbb{E}\left[D_n(k)\right]\mathbb{P}[Z=k|Z\in S_2] \notag\\
       =&\sum_{j\in S_1} \sum_{k\in S_2} \sum_{n\in S_1} \mathbb{E}\left[ D_n(j)-D_n(k)\right]\omega_{jk} \notag \\ \notag
    =&\sum_{j\in S_1}\sum_{k\in S_2} \sum_{n\in S_1}\left(\mathbb{P}\left[D_n(j)- D_n(k)=1\right]-
    \mathbb{P}\left[D_n(j)- D_n(k)=-1\right]\right)\omega_{jk} \\ \notag
    =& \sum_{j\in S_1}\sum_{k\in S_2} \sum_{l\in S_1} \sum_{m\in S_2}\left(\mathbb{P}\left[D(j)=l,D(k)=m\right]-
    \mathbb{P}\left[D(j)=m, D(k)=l\right]\right)\omega_{jk},
\end{align}
where we use $D_n=\sum_{l=1}^J D_n(l) \times Z_l$ and Assumption~\ref{ass:standardIV}(a), $\sum_{j\in S_a}\mathbb{P}[Z=j|Z\in S_a] =1$ and $\sum_{k\in S_b}\mathbb{P}[Z=k|Z\in S_b] =1$,  $\omega_{jk}=\mathbb{P}[Z=j|Z\in S_a]\mathbb{P}[Z=k|Z\in S_b]$, and 
\begin{align*}
    \sum_{j\in S_1}\sum_{k\in S_2} \sum_{l\in S_1} \sum_{m\in S_1}\left(\mathbb{P}\left[D(j)=l,D(k)=m\right]-
    \mathbb{P}\left[D(j)=m, D(k)=l\right]\right)\omega_{jk}=0.
\end{align*}
The numerator of \eqref{eq:beta2} can be written as
\begin{align}
    \Delta\mathbb{E}[Y|\tilde{Z}_1]=& \sum_{l=1}^J \mathbb{E}[Y(l)  D_l]\tilde{Z}=1] -\sum_{l=1}^J \mathbb{E}[Y(l)  D_l]\tilde{Z}=2] \\ \notag
    =& \sum_{j\in S_1} \sum_{l=1}^J \mathbb{E}\left[Y(l) D_l(j) |Z=j\right]\mathbb{P}[Z=j|Z\in S_1] - 
       \sum_{k\in S_2} \sum_{l=1}^J \mathbb{E}\left[Y(l) D_l(k) |Z=k\right]\mathbb{P}[Z=k|Z\in S_2]\\ \notag
       =& \sum_{j\in S_1} \sum_{k\in S_2} \sum_{l=1}^J \mathbb{E}\left[Y(l) (D_l(j)-D_l(k))\right]\omega_{jk}\\ \notag
       =& \sum_{j\in S_1} \sum_{k\in S_2} \sum_{l=1}^J \sum_{m \neq l}\mathbb{E}\left[Y(l)-Y(m)| D(j)=l,D(k)=m\right]\mathbb{P}[D(j)=l,D(k)=m]\omega_{jk},
\end{align}
where we use Assumption~\ref{ass:standardIV}(b) to substitute for potential outcomes $Y=\sum_{l=1}^J Y(l) \times D_l$, and Assumption~\ref{ass:standardIV}(a). 

Define $\Delta_{jklm}=\mathbb{E}\left[Y(l)-Y(m)| D(j)=l,D(k)=m\right]$ and $P_{jklm}=\mathbb{P}[D(j)=l,D(k)=m]$. Now we can write:
\begin{align}
    \sum_{l=1}^J \sum_{m \neq l}\Delta_{jklm}P_{jklm}=& 
    \sum_{l \in S_1}\sum_{m  \in S_2} \Delta_{jklm}P_{jklm} + \sum_{l \in S_2}\sum_{m  \in S_1} \Delta_{jklm}P_{jklm} + \\ \notag
    &\sum_{l \in S_1}\sum_{m \neq l \in S_1} \Delta_{jklm}P_{jklm} +
    \sum_{l \in S_2}\sum_{m \neq l \in S_2} \Delta_{jklm}P_{jklm}\\ \notag
    =& \sum_{l \in S_1}\sum_{m  \in S_2} \Delta_{jklm}P_{jklm} - \Delta_{jkml}P_{jkml} + \\ \notag
    &\sum_{l \in S_1}\sum_{m \neq l \in S_1} \Delta_{jklm}P_{jklm} +
    \sum_{l \in S_2}\sum_{m \neq l \in S_2} \Delta_{jklm}P_{jklm}.
\end{align}

Hence, according to \Cref{ass:relevance}, 
\begin{align}
    \beta = \frac{\sum_{j\in S_1} \sum_{k\in S_2}\omega_{jk} \left[ \sum_{l \in S_1}\sum_{m  \in S_2} \Delta_{jklm}P_{jklm} - \Delta_{jkml}P_{jkml} \right]+B_1 +B_2}{\sum_{j\in S_1} \sum_{k\in S_2}\omega_{jk}\left[ \sum_{l \in S_1}\sum_{m  \in S_2} P_{jklm} - P_{jkml}\right]},
\end{align}
with $B_c = \sum_{j\in S_1} \sum_{k\in S_2}\omega_{jk}\sum_{l \in S_c}\sum_{m \neq l \in S_c} \Delta_{jklm}P_{jklm}$.

\subsection{Proof Theorem~\ref{thm:clateC}}\label{A:proof_clate}
% derive equation for beta_ab
Using Assumption~\ref{ass:relevance}, extended to the case with more than two treatment clusters, the expression of the IV estimand in \eqref{eq:betaC} can be written as
\begin{align}
\mathbb{E}[\mathbf{\tilde{Z}_b}\mathbf{\tilde{D}_b}']\mathbf{\beta_b} = \mathbb{E}[\mathbf{\tilde{Z}_b}Y],
\end{align}
from which follows that the IV parameters $\beta_{ab}$ equal
\begin{align}\label{eq:betaf}
    \beta_{ab} = \frac{{\theta}_{ab}}{{\pi}_{aab}}-\sum_{c \neq a,b} \frac{{\pi}_{cab}}{{\pi}_{aab}} {\beta}_{cb}, \quad \forall a \neq b,
\end{align}
with 
\begin{align}
    {\theta}_{ab} = \mathbb{E}[Y|\tilde{Z}=a]-\mathbb{E}[Y|\tilde{Z}=b] \text{ and }
    {\pi}_{cab} =  \mathbb{E}[\tilde{D}_c|\tilde{Z}=a]-\mathbb{E}[\tilde{D}_c|\tilde{Z}=b].
\end{align}

% write in terms of potential outcomes
Write ${\pi}_{cab}$ for all $c\neq a,b$ in terms of individual treatments
\begin{align} 
    {\pi}_{cab} =&  
       \sum_{n\in S_c} \mathbb{E}[D_n|\tilde{Z}=a]-  \sum_{n\in S_c}\mathbb{E}[D_n|\tilde{Z}=b] \\
      =& \sum_{j\in S_a} \sum_{n\in S_c} \mathbb{E}\left[D_n|Z=j \right]\mathbb{P}[Z=j|Z\in S_a] - 
       \sum_{k\in S_b} \sum_{n\in S_c}\mathbb{E}\left[D_n|Z=k\right]\mathbb{P}[Z=k|Z\in S_b]. \notag
\end{align}
Substitute for potential outcomes $D_n=\sum_{l=1}^J D_n(l) \times Z_l$, and use Assumption~\ref{ass:standardIV}(a) to write 
\begin{align}
    {\pi}_{cab} =& \sum_{j\in S_a} \sum_{n\in S_c} \mathbb{E}\left[D_n(j)\right]\mathbb{P}[Z=j|Z\in S_a] - 
       \sum_{k\in S_b} \sum_{n\in S_c}\mathbb{E}\left[D_n(k)\right]\mathbb{P}[Z=k|Z\in S_b].
\end{align}
Since $\sum_{j\in S_a}\mathbb{P}[Z=j|Z\in S_a] =1$ and $\sum_{k\in S_b}\mathbb{P}[Z=k|Z\in S_b] =1$, we have 
\begin{align}\label{eq:piflows}
    {\pi}_{cab} = &\sum_{j\in S_a} \sum_{k\in S_b} \sum_{n\in S_c} \mathbb{E}\left[ D_n(j)-D_n(k)\right]\omega_{jk} \\ \notag
    =&\sum_{j\in S_a}\sum_{k\in S_b} \sum_{n\in S_c}\left(\mathbb{P}\left[D_n(j)- D_n(k)=1\right]-
    \mathbb{P}\left[D_n(j)- D_n(k)=-1\right]\right)\omega_{jk},
\end{align}
where $\omega_{jk}=\mathbb{P}[Z=j|Z\in S_a]\mathbb{P}[Z=k|Z\in S_b]$.

%apply assumptions 2-4 to remove bias
According to Assumption~\ref{ass:mono}, $\mathbb{P}[D_n(j)-D_n(k)=1]=\mathbb{P}[D_n(j)-D_n(k)=1, D_l(j)=0] =\mathbb{P}[D_n(j)-D_n(k)=1, D_l(j)=0, D_l(k)=0]$ with $l \in S_a$. Since this holds for all $l \in S_a$, it follows from Assumption~\ref{ass:irrelevance} that $\mathbb{P}[D_n(j)-D_n(k)=1]=0$ for all $n\notin S_a$, $j\in S_a$ and $k \in S_b$. This implies that ${\pi}_{cab}$ simplifies to
\begin{align}
    {\pi}_{cab}= -\sum_{j\in S_a}\sum_{k\in S_b} \sum_{n\in S_c}\mathbb{P}\left[D_n(j)- D_n(k)=-1\right]\omega_{jk}.
\end{align}
Assumption~\ref{ass:mono} together with Assumption~\ref{ass:irrelevance} impose that when $D_n(j)-D_n(k)=-1$ it must be that $D_l(j)-D_l(k)=1$ for one treatment $l \in S_a$. Hence, we can be explicit about the clusters in ${\pi}_{cab}$:
\begin{align}
    {\pi}_{cab}= -\sum_{j\in S_a}\sum_{k\in S_b} \sum_{n\in S_c}\sum_{l\in S_a}\mathbb{P}\left[D_n(j)- D_n(k)=-1,D_l(j)- D_l(k)=1\right]\omega_{jk}.
\end{align}
Assumption~\ref{ass:control} implies that $\mathbb{P}[D_n(j)-D_n(k)=-1,D_l(j)=1]=\mathbb{P}[D_n(j)-D_n(k)=-1,D_l(j)-D_l(k)=1]=0$ for all $n \notin \{S_a, S_b\}$, $j\in S_a$ and $k\in S_b$, and
\begin{align}\label{eq:offdiag}
    {\pi}_{cab}=0,  \quad \forall c \neq a,b.
\end{align}
It follows from \eqref{eq:betaf} and \eqref{eq:offdiag} that
\begin{align}\label{eq:fsrf}
     {\beta}_{ab} = \frac{{\theta}_{ab}}{{\pi}_{aab}},  \quad \forall a \neq b. 
\end{align}

% derive expression first stage
Use the expression in \eqref{eq:piflows} to write 
\begin{align}
    {\pi}_{aab} 
     =& \sum_{j\in S_a} \sum_{k\in S_b}  \sum_{l\in S_a} \left(\mathbb{P}\left[ D_l(j)- D_l(k)=1\right]
    -
   \mathbb{P}\left[D_l(j)- D_l(k)=-1\right]\right)
    \omega_{jk}.
\end{align}
Assumption~\ref{ass:mono} implies that $\mathbb{P}[D_l(j)-D_l(k)=-1]=0$ for all $j,l\in S_a$ and $k\in S_b$, so that 
\begin{align}
   {\pi}_{aab} = \sum_{j\in S_a} \sum_{k\in S_b}  \sum_{l\in S_a} \mathbb{P}\left[ D_l(j)- D_l(k)=1\right]\omega_{jk}.
\end{align}
%Since Assumption~\ref{ass:control} implies that $D(k)$ is in either $S_a$ or $S_b$, and Assumption~\ref{ass:mono} implies $D(k)$ cannot be in $S_a$, we have $D(j)=l$ and $D(k)=m$ with $m \in S_b$,
Since Assumption~\ref{ass:control} implies that when $D_l(j)- D_l(k)=1$ it must be that $D_m(j)- D_m(k)=-1$ with $m$ in $S_a$ or $S_b$, and Assumption~\ref{ass:mono} implies that $m \notin S_a$,
\begin{align}\label{eqA:clate_fs}
    {\pi}_{aab} &=\sum_{j\in S_a} \sum_{k\in S_b}  \sum_{l\in S_a} \sum_{m\in S_b}
    \mathbb{P}\left[ D_l(j)- D_l(k)=1, D_m(j)- D_m(k)=-1\right]
    \omega_{jk}.
\end{align}

% derive expression reduced form
Next, we use Assumption~\ref{ass:standardIV}(b) to substitute for potential outcomes $Y=\sum_{l=1}^J Y(l) \times D_l$, and write
\begin{align}
    {\theta}_{ab}=&\sum_{l=1}^J \mathbb{E}[Y(l)  D_l|\tilde{Z}=a] -\sum_{l=1}^J \mathbb{E}[Y(l)  D_l|\tilde{Z}=b] \\ \notag
    =& \sum_{j\in S_a} \sum_{l=1}^J \mathbb{E}\left[Y(l) D_l(j) |Z=j\right]\mathbb{P}[Z=j|Z\in S_a] - 
       \sum_{k\in S_b} \sum_{l=1}^J \mathbb{E}\left[Y(l) D_l(k) |Z=k\right]\mathbb{P}[Z=k|Z\in S_b],
    \end{align}
then use Assumption~\ref{ass:standardIV}(a) to arrive at
\begin{align}
    {\theta}_{ab}=\sum_{j\in S_a} \sum_{k\in S_b} \sum_{l=1}^J \mathbb{E}\left[Y(l) (D_l(j)-D_l(k))\right]\omega_{jk},
\end{align}
where we split the summation across treatments $l$ into clusters,
\begin{align}\label{eqA:clate_rf_expand}
   {\theta}_{ab} =& \sum_{j\in S_a} \sum_{k\in S_b} \sum_{l\in S_a} \mathbb{E}\left[Y(l) (D_l(j)-D_l(k))\right]\omega_{jk} +\\ \notag
     & \sum_{j\in S_a} \sum_{k\in S_b} \sum_{m\in S_b} \mathbb{E}\left[Y(m) (D_m(j)-D_m(k))\right]\omega_{jk} +\\ \notag
     & \sum_{j\in S_a} \sum_{k\in S_b} \sum_{n\notin S_a,S_b} \mathbb{E}\left[Y(n) (D_n(j)-D_n(k))\right]\omega_{jk}. \notag
\end{align}
For the second term in \eqref{eqA:clate_rf_expand}, Assumption~\ref{ass:mono} imposes that $\mathbb{P}[D_m(j)-D_m(k)=1]=\mathbb{P}[D_m(j)-D_m(k)=1,D_l(j)=0]=\mathbb{P}[D_m(j)-D_m(k)=1, D_l(j)=0, D_l(k)=0]$ for all $l \in S_a$. Then we have according to Assumption~\ref{ass:irrelevance} that $\mathbb{P}[D_m(j)-D_m(k)=1]=0$ for all $m \in S_b$, $j\in S_a$, and $k \in S_b$. From \eqref{eq:offdiag} follows that $D_n(j)-D_n(k)=0$ for all $n \notin \{S_a, S_b\}$. Hence, the third term in \eqref{eqA:clate_rf_expand} equals zero. Moreover, it follows that $D_m(j)-D_m(k)=-1$ if $D_l(j)- D_l(k)=1$, and that $D_l(j)- D_l(k)=1$ if $D_m(j)-D_m(k)=-1$, for all $j,l\in S_a$ and $k,m\in S_b$, and so ${\theta}_{ab}$ can be written as
\begin{align}\label{eqA:clate_rf}
    {\theta}_{ab} =&  \sum_{j \in S_a} \sum_{k \in S_b}  \sum_{l \in S_a} \sum_{m \in S_b} \mathbb{E}[{Y}(l)-{Y}(m)|D_l(j)-D_l(k)=1,D_m(j)-D_m(k)=-1] \times \\ \notag
    & \mathbb{P}\left[D_l(j)-D_l(k)=1,D_m(j)-D_m(k)=-1\right]\omega_{jk}.
\end{align}

% final expression
Combining \eqref{eqA:clate_fs} and \eqref{eqA:clate_rf} with \eqref{eq:fsrf} we have
\begin{align}
     {\beta}_{ab} =
     \sum_{j \in S_a} \sum_{k \in S_b} \sum_{l \in S_a}   \sum_{m \in S_b}
     \lambda_{jklm}\mathbb{E}[{Y}(l)-{Y}(m)| D(j)=l, D(k)=m],
\end{align}
with
\begin{align}
    \lambda_{jklm} = \frac{\omega_{jk}\mathbb{P}\left[ D(j)=l, D(k)=m\right]
    }{\sum\limits_{j\in S_a} \sum\limits_{k\in S_b} \omega_{jk} \sum\limits_{l\in S_a}   \sum\limits_{m\in S_b}
    \mathbb{P}\left[ D(j)=l, D(k)=m\right]
    },
\end{align}
where we use that $(D_l(j)- D_l(k)=1, D_m(j)- D_m(k)=-1)$ corresponds to $(D(j)=l, D(k)=m)$.
Note that with $C=2$, this proves \Cref{thm:clate}.

\subsection{Proof Theorem~\ref{thm:clate_rdd}}\label{A:clate_rdd}
The estimand in \eqref{eq:betaCx} allows for the expression $\mathbf{\beta_b^h} = \mathbb{E}_h[\mathbf{\ddot{Z}_{b,h}}\mathbf{\tilde{D}_{b}}{}']^{-1}\mathbb{E}_h[\mathbf{\ddot{Z}_{b,h}}Y]$, where $\mathbf{\ddot{Z}_{b,h}}=\mathbf{\tilde{Z}_b}-\mathcal{P}_h(\mathbf{\tilde{Z}_b}|\mathbf{X}_h)$ with linear projection $\mathcal{P}_h(\mathbf{\tilde{Z}_b}|\mathbf{X}_h)=\mathbb{E}_h[\mathbf{\tilde{Z}_b}\mathbf{X}_h']\mathbb{E}_h[\tilde{X}_{h}\mathbf{X}_h']^{-1}\mathbf{X}_{h}$ and $\mathbb{E}_h[(\mathbf{\tilde{Z}_b}-\mathcal{P}_h(\mathbf{\tilde{Z}_b}|\mathbf{X}_h))]=0$ since we include an intercept.
Consider row $a$ and column $c \neq b$ from the matrix $\mathbb{E}_h[\mathbf{\ddot{Z}_{b,h}}\mathbf{\tilde{D}_{b}}{}']$:
\begin{align}
    \mathbb{E}_h[\ddot{Z}_{a,h}\tilde{D}_{c}] =& \sum_{j \in S_a}\sum_{k \in S_b} \mathbb{P}_h[f_{jk}=1]\mathbb{E}_h[\ddot{Z}_{a,h}\tilde{D}_{c}|f_{jk}=1]\\ \notag
    =& \sum_{j \in S_a}\sum_{k \in S_b} \mathbb{P}_h[f_{jk}=1]\mathbb{E}_h[\ddot{Z}_{a,h}\mathbb{E}_h[\tilde{D}_{c}|S,f_{jk}=1]|f_{jk}=1],
\end{align}
where we use that the polynomial is flexible enough so that we can write the projection residual per $f_{jk}$; $\ddot{Z}_{a,h}=\mathbbm{1}[S \geq 0]-\mathcal{P}_h(\mathbbm{1}[S \geq 0]|\mathbf{B}_h,f_{jk}=1)$ conditional on $f_{jk}=1$. Since the fixed effects are constant with $f_{jk}=1$ and the polynomial is a function of $S$, $\ddot{Z}_{a,h}$ is a measurable function of \(S\) and we can apply iterated expectations.

Under \Cref{ass:RD}, we can write
\begin{align}
    \lim_{h \downarrow 0} \mathbb{E}_h[\tilde{D}_{c}|S,f_{jk}=1] = \lim_{v \uparrow 0} \mathbb{E}[\tilde{D}_{c}|S=v,f_{jk}=1]+\mathbbm{1}[S \geq 0]\Delta_{jk}^c,
\end{align}
where we define $\Delta_{jk}^c = \lim_{v \downarrow 0} \mathbb{E}[\tilde{D}_{c}|S=v,f_{jk}=1]-\lim_{v \uparrow 0} \mathbb{E}[\tilde{D}_{c}|S=v,f_{jk}=1]$. Therefore,
\begin{align}
    \lim_{h \downarrow 0} \mathbb{E}_h[\ddot{Z}_{a,h}\mathbb{E}_h[\tilde{D}_{c}|S,f_{jk}=1]|f_{jk}=1] =& \lim_{v \uparrow 0} \mathbb{E}[\tilde{D}_{c}|S=v,f_{jk}=1]\lim_{h \downarrow 0}\mathbb{E}_h[\ddot{Z}_{a,h}|f_{jk}=1]+\notag\\&\Delta_{jk}^c\lim_{h \downarrow 0}\mathbb{E}_h[\ddot{Z}_{a,h}\mathbbm{1}[S \geq 0]|f_{jk}=1]= \Delta_{jk}^c \kappa,
\end{align}
where we use that $\mathbb{E}_h[\ddot{Z}_{a,h}|f_{jk}=1]=0$ by the projection property, and that $\lim_{h \downarrow 0}\mathbb{E}_h[\ddot{Z}_{a,h}\mathbbm{1}[S \geq 0]|f_{jk}=1]=\kappa$ as the conditional distribution of $S/h$ within $|S|\le h$ converges to the same uniform distribution for every $(j,k)$ as $h\to0$ if $S$ has a density that is continuous and strictly positive at $S=0$.

It follows that $\lim_{h \downarrow 0} \mathbb{E}_h[\ddot{Z}_{a,h}\tilde{D}_{c}] =  \kappa \sum_{j \in S_a}\sum_{k \in S_b}  \omega_{jk} \Delta_{jk}^c$, where $\omega_{jk}=\lim_{h \downarrow 0}\mathbb{P}_h[f_{jk}=1]$. We use that $\tilde{D}_{c} = \sum_{n \in S_c} D_n$ to write
\begin{align}
    \Delta_{jk}^c =& \sum_{n \in S_c} \lim_{v \downarrow 0} \mathbb{E}[{D}_{n}|S=v,f_{jk}=1]-\lim_{v \uparrow 0} \mathbb{E}[{D}_{n}|S=v,f_{jk}=1] \\ \notag
    =& \sum_{n \in S_c} \lim_{v \downarrow 0} \mathbb{E}[{D}_{n}|Z=j,S=v,f_{jk}=1]-\lim_{v \uparrow 0} \mathbb{E}[{D}_{n}|Z=k,S=v,f_{jk}=1]\\ \notag
    =& \sum_{n \in S_c} \mathbb{E}[{D}_{n}(j)-{D}_{n}(k)|S=0,f_{jk}=1],
\end{align}
where we use the definition of the instruments and \Cref{ass:RD}. Similarly, we have
\begin{align}
    \lim_{h \downarrow 0} \mathbb{E}_h[\ddot{Z}_{a,h}Y] =  \kappa \sum_{j \in S_a}\sum_{k \in S_b} \omega_{jk} \sum_{n=1}^J\mathbb{E}[Y(n)({D}_n(j)-{D}_n(k))|S=0,f_{jk}=1],
\end{align}
and since the $\kappa$ occurs in the numerator and denominator it cancels out. Now the result follows from \Cref{thm:clateC}.

\section{Additional theoretical results}
\label{A:Weights}

\Cref{thm:clateC} shows that the TSLS estimand identifies a weighted average of LATEs using only data at the clustered treatment level. When data at the unclustered treatment level is available, this Appendix provides three strategies for providing additional insights into the components of the weighted average. First, we show that the weights of the treatment effects are partially identified. Second, we consider two sets of additional assumptions which allow for the identification of both the weights and the LATEs in the weighted averages. We provide testable conditions for the additional assumptions. 

\subsection{Partial identification of the treatment effect weights}

The contribution of each component $\lambda_{jklm}\Delta_{jklm}$ to $\beta_{ab}$ is determined jointly by the LATE $\Delta_{jklm}$ and its weight $\lambda_{jklm}$. These weights represent the normalized proportion of the response types corresponding to each LATE. From a policy perspective, these weights capture how policy-induced encouragement affects treatment take up. For instance, the weights illustrate how enrollment into specific non-STEM and STEM fields respond to a policy promoting female enrollment in STEM.

In the general case in which both $S_a$ and $S_b$ contain multiple treatments, these weights cannot be point identified. In this case, individuals observed to take a treatment from $S_a$ if $Z \in S_a$ could have taken any treatment in $S_b$ if $Z \in S_b$, and the other way around.
\begin{corollary}[Partial identification treatment effect weights]\label{cor:weights}\ \\
Under Assumptions~\ref{ass:standardIV}-\ref{ass:control}, it holds for $\lambda_{jklm}$ in \eqref{eq:weight} with $j,l \in S_a$ and $k,m \in S_b$ that
\begin{align}\label{eq:weights}
 \frac{\max(0,-\pi_{mjk}-\sum_{n \in \{S_a\setminus l\}}\pi_{njk}) \omega_{jk}}{   \sum\limits_{j\in S_a}\sum\limits_{k\in S_b}\omega_{jk}\sum\limits_{l\in S_a}  \pi_{ljk}}  \leq\lambda_{jklm}\leq  \frac{\min(\pi_{ljk},-\pi_{mjk}) \omega_{jk}}{   \sum\limits_{j\in S_a}\sum\limits_{k\in S_b}\omega_{jk}\sum\limits_{l\in S_a}  \pi_{ljk}},
\end{align}
where $\pi_{ljk}=\mathbb{E}[D_l|Z=j]-\mathbb{E}[D_l|Z=k]=\mathbb{P}[D(j)=l,D(k)\neq l]-\mathbb{P}[D(j)\neq l,D(k)=l]$,
and $\omega_{jk}$ as defined in \eqref{eq:probz}.
\end{corollary}

The proof is deferred to Appendix~\ref{A:weights}. Corollary~\ref{cor:weights} provides individual bounds on each weight. These bounds do not jointly characterize the feasible weights, since extreme values for different weights may not be simultaneously attainable. 

\Cref{cor:weights} shows that the proportion of types $(D(j)=l,D(k)=m)$ is upper bounded by both the total proportion of types $(D(j)=l,D(k)\in S_b)$ and the total proportion of types $(D(j)\in S_a, D(k)=m)$. A lower bound is constructed by the difference between the total proportion of types $(D(j)\in S_a, D(k)=m)$ and the total proportion of types $(D(j) \in \{S_a \setminus l\}, D(k)\in S_b)$. Since $\pi_{ljk}$ and $\omega_{jk}$ can be identified from unclustered treatment data, the bounds can be estimated when this data is observed. 

% explain corollary with special case in which weights are point identified
When either $S_a$ or $S_b$ contains only one treatment, the bounds in \Cref{cor:weights} boil down to point identification. If $S_a=\{j\}$, the proportion of types with $(D(j)\in S_a, D(k)=m)$ coincides with $(D(j)=j, D(k)=m)$, and if $S_b=\{k\}$, the proportion of types with $(D(j)=l, D(k)\in S_b)$ coincides with $(D(j)=l, D(k)=k)$. This yields the following expressions with $S_a=\{j\}$ and $S_b=\{k\}$, respectively:
\begin{align} \label{eq:weightspoint}
  \hspace{-.04in} \lambda_{jkjm}= \frac{-\pi_{mjk} \mathbb{P}[Z=k|Z\in S_b]}{   \sum\limits_{k\in S_b}\mathbb{P}[Z=k|Z\in S_b]  \pi_{jjk}} \text{ and } \lambda_{jklk}= \frac{\pi_{ljk} \mathbb{P}[Z=j|Z\in S_a]}{   \sum\limits_{j\in S_a}\mathbb{P}[Z=j|Z\in S_a]\sum\limits_{l\in S_a}  \pi_{ljk}}. 
\end{align}

\subsection{Full identification with heterogeneity restrictions}

In addition to the treatment weights $\lambda_{jklm}$, we aim to identify the corresponding LATEs $\Delta_{jklm}$. To achieve this, we specify the estimand that targets the LATEs $\Delta_{jkjk}$ for all $j$ comparing to treatment $k$:
\begin{align}\label{eq:alpha}
    \mathbf{\alpha_k} =&(\alpha_{0k},\alpha_{1k},\dots,\alpha_{k-1,k},\alpha_{k+1,k},\dots,\alpha_{Jk})'
    = \mathbb{E}[\mathbf{{Z_k}}\mathbf{{D_k}}']^{-1}\mathbb{E}[\mathbf{{Z_k}}Y],
\end{align}
which equals $\mathbf{\beta_b}$ if $S_j=\{j\}$ for $j=1,\dots,J$ with $S_b=\{k\}$. The corollary below shows that $\alpha_{jk}=\Delta_{jkjk}$ when we impose  restrictions on treatment effect heterogeneity.  
\begin{corollary}[Treatment effect heterogeneity restrictions]\label{cor:homo}\ \\
Under Assumptions~\ref{ass:standardIV}-\ref{ass:control} and the following two conditions 
\begin{enumerate}
    \item If $j \in S_a$,  $\Delta_{jklm}=\Delta_{lklk}-\Delta_{mkmk}$ for $l \in S_a$ and $k,m \in S_b$ for all $a \neq b$,
    \item If $j \in S_b$, $\Delta_{jklm}=\Delta_{lklk}-\Delta_{mkmk}$ for $k \in S_b$ and all $l,m \in S$,
\end{enumerate}
it holds that
\begin{align}\label{eq:wa_beta}
    {\beta}_{ab} =&
    \frac{\sum_{j \in S_a}\sum_{k \in S_b} \sum_{l \in S_a}\Delta_{lklk}\pi_{ljk}\omega_{jk}}{\sum_{j \in S_a}\sum_{k \in S_b} \sum_{l \in S_a}\pi_{ljk}\omega_{jk}}- \frac{\sum_{j \in S_a}\sum_{k \in S_b}\sum_{m \in S_b}\Delta_{mkmk}\pi_{mjk}\omega_{jk}}{\sum_{j \in S_a}\sum_{k \in S_b} \sum_{m \in S_b}\pi_{mjk}\omega_{jk}},\\
    \alpha_{jk}=&\Delta_{jkjk}, \text{ with }  j \in S_a \text{ and } k \in S_b.
\end{align}
\end{corollary}
The proof is deferred to Appendix~\ref{A:homo}. Assumptions~\ref{ass:mono} to \ref{ass:control} allow for response types $(D(j)\in S_a, D(k)\in S_b)$ with $j\in S_a$ and $k \in S_b$. To identify $\alpha_{jk}$, we require restrictions on the effects between treatments across $S_a$ and $S_b$ for all $a \neq b$ that correspond to these responses (condition 1). Since the responses to instrument values within $S_b$ are not restricted, we additionally require restrictions on the treatment effects that correspond to these responses (condition 2). In case both conditions hold, the weighted average of $\alpha_{jk}$ in \eqref{eq:wa_beta} equals ${\beta}_{ab}$. Hence, the conditions are violated if the equality in \eqref{eq:wa_beta} does not hold, and \eqref{eq:wa_beta} thus serves as a necessary condition for the restrictions on treatment effect heterogeneity. 

The restrictions in \Cref{cor:homo} are substantially stronger than those required for \Cref{thm:clate}. Treatment effect heterogeneity is not allowed to be systematically related to individuals' response types. A sufficient condition is mean independence of pairwise treatment effects from the potential treatment choices. This allows expected outcome levels to differ across response types, but requires the relative effects of fields of study not to depend on how students' enrollment responds to field offers. Such restrictions may be more plausible when students have limited information about their field-specific gains, so that sorting on those gains is less likely and choices instead reflect access constraints or geographic considerations.

\subsection{Full identification under response type restrictions}

Alternatively, the LATEs can be identified by ruling out additional response types. Consider Assumptions \ref{ass:mono} to \ref{ass:control}, with each cluster only including one treatment, such that $S_j=\{j\}$ for $j=1,\dots,J$. These assumptions only allow for response types with $D(j) \neq D(k)$ that satisfy $(D(j)=j,D(k)=k)$ with $j \in S_a$ and $k \in S_b$. Thus, each instrument can only induce take up of its intended treatment. 
\begin{corollary}[Response type restrictions]\label{cor:alpha}\ \\
Under \Cref{ass:standardIV}, and \Cref{ass:relevance}-\ref{ass:control} at the unclustered treatment level, it holds that
\begin{align}\label{eq:singleshift}
    \beta_{ab}=&\sum_{j \in S_a} \sum_{k \in S_b} \lambda_{jkjk}\Delta_{jkjk}, \quad \lambda_{jkjk}=\frac{\pi_{jjk}\omega_{jk}}{\sum_{j \in S_a} \sum_{k \in S_b}\pi_{jjk}\omega_{jk}},\\
    \alpha_{jk} =& \Delta_{jkjk}, \text{ with }  j \in S_a \text{ and } k \in S_b.
\end{align}
\end{corollary}
The proof is deferred to \Cref{A:alpha}. The assumptions in \Cref{cor:alpha} imply that $\phi_{jjk}^A\geq0$ and $\phi_{ljk}^A=0$ for all $l \neq j,k$. 
Similar to \Cref{prop:nec_cond} at the level of treatment clusters, these $\phi_{ljk}^A$ pose testable conditions for Assumptions \ref{ass:mono} to \ref{ass:control} at the unclustered treatment level. 

The conditions at the unclustered level are more demanding than their cluster-level counterparts: if they hold at the unclustered level, the corresponding cluster-level conditions also hold, but not vice versa. Similar conditions for identifying LATEs between unclustered treatments are discussed by \citet{behaghel2013robustness} and \citet{bhuller20242sls}. In the context of fields of study, \citet{kirkeboen2016field} invoke Assumptions~\ref{ass:relevance} to \ref{ass:control} at the unclustered level and subsequently aggregate fields into ten clusters in their empirical application. \Cref{cor:alpha} shows that the resulting treatment cluster estimand identifies a weighted average of LATEs between individual fields.

\subsection{Proof Corollary~\ref{cor:weights}}\label{A:weights}
From \Cref{ass:standardIV}(a) follows that for all $l\neq k$
\begin{align}
    \pi_{ljk}=\mathbb{E}[D_l(j)-D_l(k)]=\mathbb{P}[D_l(j)-D_l(k)=1]-\mathbb{P}[D_l(j)-D_l(k)=-1].
\end{align}

From \Cref{thm:clateC} we have that
\begin{align}
    \lambda_{jklm} = \frac{\omega_{jk}\mathbb{P}\left[D_l(j)-D_l(k)=1,D_m(j)-D_m(k)=-1\right]
    }{\sum\limits_{j\in S_a} \sum\limits_{k\in S_b} \omega_{jk} \sum\limits_{l\in S_a}   \sum\limits_{m\in S_b}
    \mathbb{P}\left[D_l(j)-D_l(k)=1,D_m(j)-D_m(k)=-1\right]
    },
\end{align}
where $(D_l(j)- D_l(k)=1, D_m(j)- D_m(k)=-1)$ corresponds to $(D(j)=l, D(k)=m)$.
The denominator can be written as 
\begin{align}\label{eq:w_denom}
    \sum\limits_{j\in S_a} \sum\limits_{k\in S_b} \omega_{jk}\sum\limits_{l\in S_a}   \sum\limits_{m\in S_b}
    \mathbb{P}\left[D_l(j)-D_l(k)=1,D_m(j)-D_m(k)=-1\right]
    =\sum\limits_{j\in S_a}\sum\limits_{k\in S_b} \omega_{jk}\sum\limits_{l\in S_a} \pi_{ljk}.
\end{align}
% upper and lower bound
For the probability in the numerator we have that
\begin{align}
\max(0,\mathbb{P}[D_m(j)-D_m(k)=-1]-\sum_{n \in \{S_a\setminus l\}}\mathbb{P}[D_n(j)-D_n(k)=1])\\ \notag
    \leq \mathbb{P}\left[D_l(j)-D_l(k)=1,D_m(j)-D_m(k)=-1\right] \leq \\ \notag
    \min\left(\mathbb{P}[D_l(j)-D_l(k)=1],\mathbb{P}[D_m(j)-D_m(k)=-1]\right).
\end{align}

From Assumption~\ref{ass:mono} it follows that $\mathbb{P}[D_l(j)-D_l(k)=-1]=0$ for all $l \in S_a$, and from Assumptions~\ref{ass:mono} and \ref{ass:irrelevance} that $\mathbb{P}[D_l(j)-D_l(k)=1]=0$ for all $l \in S_b$. Therefore, we can write
\begin{align}
    \max(0,-\pi_{mjk}-\sum_{n \in \{S_a\setminus l\}}\pi_{njk})\\ \notag
    \leq \mathbb{P}\left[D_l(j)-D_l(k)=1,D_m(j)-D_m(k)=-1\right] \leq
    \min\left(\pi_{ljk},-\pi_{mjk}\right).
\end{align}

\subsection{Proof Corollary~\ref{cor:homo}}\label{A:homo}
From \Cref{thm:clateC} follows that 
\begin{align}
    {\beta}_{ab} = \sum_{j \in S_a}\sum_{k \in S_b}\sum_{l \in S_a}\sum_{m \in S_b}\lambda_{jklm}\Delta_{jklm}, \label{eq:betahomo}
\end{align}
where we use additional condition 1 that $\Delta_{jklm}=\Delta_{lklk}-\Delta_{mkmk}$ for $j,l \in S_a$ and $k,m \in S_b$ with $a \neq b$ to write
\begin{align}
    {\beta}_{ab} &=
     \sum_{j \in S_a}\sum_{k \in S_b} \left(\sum_{l \in S_a}\Delta_{lklk}\sum_{m \in S_b} \lambda_{jklm}- \sum_{m \in S_b}\Delta_{mkmk}\sum_{l \in S_a} \lambda_{jklm}
    \right),
\end{align}
in which, using Assumptions~\ref{ass:standardIV}-\ref{ass:control} and \eqref{eq:w_denom},
\begin{align}
    \sum_{m \in S_b} \lambda_{jklm} = \frac{\omega_{jk}\pi_{ljk}}{\sum\limits_{j\in S_a}\sum\limits_{k\in S_b} \omega_{jk}\sum\limits_{l\in S_a} \pi_{ljk}},\quad
    \sum_{l \in S_a} \lambda_{jklm} = \frac{\omega_{jk}\pi_{mjk}}{\sum\limits_{j\in S_a}\sum\limits_{k\in S_b} \omega_{jk}\sum\limits_{m\in S_b} \pi_{mjk}}
\end{align}
where we use that $\pi_{ljk}=\mathbb{P}[D_l(j)-D_l(k)=1]$ and $\pi_{mjk}=-\mathbb{P}[D_m(j)-D_m(k)=-1]$ with $l \in S_a$ and $m \in S_b$. Therefore,
\begin{align}
  {\beta}_{ab} &=  \frac{\sum_{j \in S_a}\sum_{k \in S_b} \sum_{l \in S_a}\Delta_{lklk}\pi_{ljk}\omega_{jk}}{\sum_{j \in S_a}\sum_{k \in S_b} \sum_{l \in S_a}\pi_{ljk}\omega_{jk}}- \frac{\sum_{j \in S_a}\sum_{k \in S_b}\sum_{m \in S_b}\Delta_{mkmk}\pi_{mjk}\omega_{jk}}{\sum_{j \in S_a}\sum_{k \in S_b} \sum_{m \in S_b}\pi_{mjk}\omega_{jk}}.
\end{align}

% derive alpha
Using Assumption~\ref{ass:relevance}, extended to the case with more than two treatments at the unclustered level, the expression of the IV estimand in \eqref{eq:alpha} can be written as
\begin{align}
\mathbb{E}[\mathbf{{Z}_k}\mathbf{{D}_k}']\mathbf{\alpha_k} = \mathbb{E}[\mathbf{{Z}_k}Y],
\end{align}
from which follows that 
\begin{align}\label{eq:alphaf}
    \theta_{jk} = \sum_{l \neq k} {\pi}_{ljk} {\alpha}_{lk}, \quad \forall j \neq k,
\end{align}
with respectively 
\begin{align}
    {\theta}_{jk} = \mathbb{E}[Y|{Z}=j]-\mathbb{E}[Y|{Z}=k] \text{ and }
    {\pi}_{ljk} =  \mathbb{E}[{D}_l|{Z}=j]-\mathbb{E}[{D}_l|{Z}=k].
\end{align}

Use Assumption~\ref{ass:standardIV}(b) to substitute for potential outcomes $Y=\sum_{l=1}^J Y(l) \times D_l$, subsequently using Assumption~\ref{ass:standardIV}(a),
\begin{align}
    \theta_{jk} =& \sum_{l=1}^J\left(\mathbb{E}[Y(l)D_l|{Z}=j]-\mathbb{E}[Y(l)D_l|{Z}=k]\right)\\ \notag
    =& \sum_{l=1}^J\left(\mathbb{E}[Y(l)D_l(j)|{Z}=j]-\mathbb{E}[Y(l)D_l(k)|{Z}=k]\right)\\ \notag
    =& \sum_{l=1}^J\mathbb{E}[Y(l)(D_l(j)-D_l(k))].
\end{align}
This can be further rewritten to
\begin{align} \label{eq:theta_l}
    \theta_{jk} =& \sum_{l=1}^J\mathbb{E}[Y(l)|D_l(j)-D_l(k)=1]\mathbb{P}[D_l(j)-D_l(k)=1]-\\ \notag &\sum_{m=1}^J\mathbb{E}[Y(m)|D_m(j)-D_m(k)=-1]\mathbb{P}[D_m(j)-D_m(k)=-1]\\ \notag
    =& \sum_{l=1}^J\sum_{m \neq l}\mathbb{E}[Y(l)|D_l(j)-D_l(k)=1,D_m(j)-D_m(k)=-1]\times\\ \notag &\mathbb{P}[D_l(j)-D_l(k)=1,D_m(j)-D_m(k)=-1]-\\ \notag &\sum_{m=1}^J\sum_{l \neq m}\mathbb{E}[Y(m)|D_l(j)-D_l(k)=1,D_m(j)-D_m(k)=-1]\times\\ \notag &\mathbb{P}[D_l(j)-D_l(k)=1,D_m(j)-D_m(k)=-1]\\ \notag
    =&\sum_{l=1}^J\sum_{m \neq l}\Delta_{jklm}\mathbb{P}[D_l(j)-D_l(k)=1,D_m(j)-D_m(k)=-1]. 
\end{align}
%Use additional condition 2 that $\Delta_{jklm}=\Delta_{lklk}-\Delta_{mkmk}$. 
Recall that $k \in S_b$. If $j \in S_a$, we have according to Assumptions \ref{ass:mono} to \ref{ass:control} that $l \in S_a$ and $m \in S_b$. In this case, it follows from additional condition 1 that $\Delta_{jklm}=\Delta_{lklk}-\Delta_{mkmk}$. If $j \in S_b$, it follows from additional condition 2 that $\Delta_{jklm}=\Delta_{lklk}-\Delta_{mkmk}$. So we can rewrite \eqref{eq:theta_l} to
\begin{align}
    \theta_{jk}=&\sum_{l=1}^J \sum_{m \neq l} (\Delta_{lklk}-\Delta_{mkmk})\mathbb{P}[D_l(j)-D_l(k)=1,D_m(j)-D_m(k)=-1] \label{eq:theta_case3}\\ \notag
    =&\sum_{l=1}^J  \Delta_{lklk}\sum_{m \neq l }\mathbb{P}[D_l(j)-D_l(k)=1,D_m(j)-D_m(k)=-1]-\\& \notag
    \sum_{m=1}^J \Delta_{mkmk}\sum_{l \neq m} \mathbb{P}[D_l(j)-D_l(k)=1,D_m(j)-D_m(k)=-1]\\ \notag
    =&\sum_{l=1}^J \Delta_{lklk}\mathbb{P}[D_l(j)-D_l(k)=1]-
    \sum_{m=1}^J \Delta_{mkmk}\mathbb{P}[D_m(j)-D_m(k)=-1]\\ \notag
    =&\sum_{l \neq k} \Delta_{lklk} \pi_{ljk}. 
\end{align}

Combining \eqref{eq:alphaf} with the result that $\theta_{jk}=\sum_{l \neq k} \Delta_{lklk} \pi_{ljk}$ for all $j \neq k$ with $k \in S_b$ we have the homogeneous system of equations
\begin{align}
     \sum_{l\neq k}\left(\Delta_{lklk}-\alpha_{lk}\right)\pi_{ljk}=0,\quad  \forall  j \neq k,
\end{align}
which has the unique solution $\alpha_{lk}=\Delta_{lklk}$ for all $l \neq k$ if the matrix $\Pi_k$ with elements $\pi_{ljk}$ has full rank. Since $\mathbb{E}\left[\mathbf{D}_{k}\mathbf{Z}_{k}'\right]$ has full rank according to Assumption~\ref{ass:relevance}, extended to the case with more than two treatments at the unclustered level, and $\mathbb{E}\left[\mathbf{Z}_{k}\mathbf{Z}_{k}'\right]$ has full rank, 
$\mathbb{E}\left[\mathbf{D}_{k}\mathbf{Z}_{k}'\right]\mathbb{E}\left[\mathbf{Z}_{k}\mathbf{Z}_{k}'\right]^{-1}=[\delta_k, \Pi_k]$ has full rank with $\delta_k=(\delta_{1k},\dots,\delta_{k-1,k},\delta_{k+1,k},\dots,\delta_{J,k})'$. Therefore, $\Pi_k$ has full rank with elements $\pi_{ljk}$ for all $l\neq k$ and $j \neq k$. 

\subsection{Proof Corollary~\ref{cor:alpha}}\label{A:alpha}

Using \Cref{ass:mono} at the individual treatment level we can write $\mathbb{P}[D_j(j)-D_j(k)=-1]=0$, and with $l\neq j$, $\mathbb{P}[D_l(j)-D_l(k)=1]=\mathbb{P}[D_l(j)-D_l(k)=1,D_j(j)=0]=\mathbb{P}[D_l(j)-D_l(k)=1,D_j(j)=0,D_j(k)=0]$, which is equal to zero under \Cref{ass:irrelevance} at the individual treatment level. Under \Cref{ass:control} at the individual treatment level, we have that $D_m(k)=0$ if $D_j(j)=1$ with $m\neq j,k$. Hence, $\mathbb{P}\left[D_j(j)-D_j(k)=1,D_m(j)-D_m(k)=-1\right]=0$ with $m\neq j,k$. It follows that $\mathbb{P}\left[D_l(j)-D_l(k)=1,D_m(j)-D_m(k)=-1\right]=0$ if $l \neq j$ and/or $m \neq k$, and therefore $\lambda_{jklm}=0$ if $l \neq j$ and/or $m \neq k$. Substituting this into \Cref{thm:clateC}, we have
\begin{align}
     \beta_{ab}=\sum_{j \in S_a} \sum_{k \in S_b} \lambda_{jkjk}\Delta_{jkjk}, \quad \lambda_{jkjk}=\frac{\pi_{jjk}\omega_{jk}}{\sum_{j \in S_a} \sum_{k \in S_b}\pi_{jjk}\omega_{jk}},
\end{align}
where we use that from \Cref{ass:standardIV}(a) follows that
\begin{align}
    \pi_{jjk}=&\mathbb{E}[D_j(j)-D_j(k)]=\mathbb{P}[D_j(j)-D_j(k)=1]-\mathbb{P}[D_j(j)-D_j(k)=-1]\\ \notag
    =&\mathbb{P}[D_j(j)-D_j(k)=1]=\mathbb{P}\left[D_j(j)-D_j(k)=1,D_k(j)-D_k(k)=-1\right],
\end{align}
using subsequently Assumptions~\ref{ass:mono} and \ref{ass:control} at the individual treatment level.

If Assumptions \ref{ass:mono}-\ref{ass:control} hold on the individual treatment level, we have that $\alpha_{jk}=\beta_{ab}=\lambda_{jkjk}\Delta_{jkjk}=\Delta_{jkjk}$ if $S_j=\{j\}$ for all $j$.

\section{Additional details empirical application}\label{A:detailsapplication}

\subsection{Deferred acceptance in Portugal}\label{A:context}

% paragraph about what dges does: ranking students
Each course has a quota ($q$) of applicants that can be admitted, which are set in agreement between the institutions and the Minist\'{e}rio da Ci\^{e}ncia, Tecnologia e Ensino Superior (MCTES; Ministry of Science, Technology and Higher Education) and are relatively stable across years. Students submit their rankings to the Dire\c{c}\~{a}o-Geral do Ensino Superior (DGES; Directorate General for Higher Education), which runs the centralized admission process. DGES ranks for each course the applicants by their application score. This score is a weighted average of an applicant's high school and national exam grades. Since each course weighs the high school grades and the national exams differently, the application scores are applicant-course specific (Article 33 until 36 of \cite*{dges2019}).

% paragraph about assignment mechanism
The applicants are matched to courses using an iterative process (Article 37 of \cite*{dges2019}), similar to the college-proposing Gale-Shapley algorithm \citep*{gale1962college}:  
\begin{itemize}
\item Step 1: Each course makes offers to the first $q$ applicants on its ranked list. Each applicant keeps the best offer according to her submitted ranking and rejects the rest.
\item Step $s \geq 2$: Any course rejected in step $s-1$ by $n \geq 1$ applicants proposes to the next $n$ applicants on its list. Each applicant keeps the best offer according to her submitted ranking among all the proposals received in step $s$ (including the one retained in step $s-1$) and rejects the rest.
\end{itemize}
The algorithm terminates when the allocation is not updated, and matches each course to the set of $q$ most preferred applicants who did not reject it. 

One to two weeks before the start of the academic year, each applicant receives the offer from her matched course. Within that week, applicants have to accept the offer, participate in a second cycle, or withdraw from the application process. The slots that have remained or were not accepted after the first cycle are then allocated to the applicants who participate in the second cycle according to the same algorithm. The process concludes with an identical third cycle. 
%Second and third cycle applicants must submit a new ranking of up to six courses among the remaining courses only. 

\subsection{Data}\label{A:data}

% first data set
We use data from two different sources. First, the higher education application data from DGES contains for each applicant the submitted course ranking, their application scores, the matched course, and the course cutoffs, from 2008-09 to 2019-20. We consider for each student the first cycle from the earliest observed application in which the student also enrolls in the first year. We refer to students who applied in that cycle within the same year as a cohort. 

% second data set
The second data set has the student outcome data from the Dire\c{c}\~{a}o-Geral de Estat\'{i}sticas da Educa\c{c}\~{a}o e Ci\^{e}ncia (DGEEC; Directorate for Education Statistics). For each enrolled student, this data set contains the enrolled course, whether the student is active in the first, second, or third year of that course, and the achieved number of ECTS study credits in that course in the previous year, from 2013-14 to 2019-20. Our main treatment and outcome variable are enrollment and ECTS credits in the second year of a course. In case a student is active in the second year of a course across multiple years, these variables will refer to the first observed enrollment in the second year of a course.  

The student outcome data can be merged to the application data if we observe a student's enrollment in the first year. Hence, the first cohort for which we can combine the data sets is 2013-14. We take 2016-17 as the final cohort, so that we can use course enrollment in the second year up to 2018-19, allowing for a one-year delay. This results in the final data set with four cohorts from 2013-14 unto 2016-17. 

\begin{table}
\footnotesize
\begin{centering}
    \caption{Descriptive statistics for estimation sample}
    \label{tbl:DescrStats}
\renewcommand{\arraystretch}{0.70}
\begin{tabular}{l l | l l l l} 
\toprule\toprule
\multicolumn{6}{c}{Panel A: Means and standard deviations} \\
\multicolumn{2}{l}{} & Mean && Std. dev.& \\
[.5em]
\multicolumn{2}{l}{Running variable (re-centered)} & 8.55 &&18.12& \\
\multicolumn{2}{l}{$\mathbbm{1}{(\text{running variable}>0)}$} &0.68&&& \\
\multicolumn{2}{l}{No. of courses ranked} &5.46&&0.99&  \\
\multicolumn{2}{l}{$\mathbbm{1}{(\text{top two fields} = \{ r_1, r_2\}) }$} &0.88&&&  \\
\multicolumn{2}{l}{$\mathbbm{1}{(D \neq Z)}$ in year 2} &0.08&&& \\
\multicolumn{2}{l}{No. of ECTS in year 2} &54.19&&13.52&\\
\multicolumn{2}{l}{Observations (cohorts)} &49272 (4)&&&\\
\midrule \\
[-1.5em]
\multicolumn{6}{c}{Panel B: Largest preferred fields per next-best} \\
Next-best field & Total obs. & \multicolumn{2}{c}{Largest pref. field} &  \multicolumn{2}{c}{2nd largest pref. field} \\
&  & Name& Obs. & Name & Obs. \\
[.5em]
Agriculture, forestry and fishery	&	363	&	Veterinary	&	153	&	Life sciences	&	64	\\
Architecture and building	&	2063	&	Engineering 	&	1004	&	Arts	&	762	\\
Arts	&	1735	&	Architecture	&	578	&	Humanities	&	271	\\
Business and administration	&	7432	&	Soc. and beh. sci.	&	3001	&	Personal services	&	815	\\
Computing	&	1755	&	Engineering 	&	1366	&	Bus. and adm. &	292	\\
Engineering and eng. trades	&	5214	&	Health	&	1901	&	Life sciences	&	1001	\\
Environmental protection	&	982	&	Engineering &	362	&	Health	&	248	\\
Health	&	2855	&	Life sciences	&	1052	&	Engineering &	725	\\
Humanities	&	2875	&	Soc. and beh. sci.	&	883	&	Bus. and adm.	&	549	\\
Journalism and information	&	1313	&	Bus. and adm.	&	338	&	Soc. and beh. sci.	&	287	\\
Law	&	1379	&	Bus. and adm.	&	473	&	Soc. and beh. sci.	&	456	\\
Life sciences	&	4340	&	Health	&	2341	&	Engineering	&	1157	\\
Manufacturing and processing	&	1102	&	Engineering	&	680	&	Health	&	101	\\
Mathematics and statistics	&	926	&	Engineering &	389	&	Bus. and adm.	&	172	\\
Personal services	&	1813	&	Bus. and adm.	&	876	&	Soc. and beh. sci.	&	202	\\
Physical sciences	&	2672	&	Engineering	&	1132	&	Life sciences	&	607	\\
Security services	&	68	&	Personal services	&	40	&	Health	&	28	\\
Social and behavioural science	&	7014	&	Bus. and adm.	&	2500	&	Law	&	1173	\\
Social services	&	1264	&	Soc. and beh. sci.	&	408	&	Teacher	&	272	\\
Teacher train. and education sci.	&	1370	&	Soc. and beh. sci.	&	510	&	Social services	&	210	\\
Transport services	&	21	&	Engineering	&	21	&	$-$	&	$-$	\\
Veterinary	&	716	&	Health	&	587	&	Life sciences	&	59	\\
      \bottomrule \bottomrule
    \end{tabular}
    \begin{tablenotes}
	\footnotesize
	\item[] Notes: Panel A shows descriptive statistics of our estimation sample  of 49,272 applicants from 4 cohorts. Panel B provides an overview of the number of applicants per next-best and preferred field combination. 
\end{tablenotes}
    \end{centering}
\end{table}

To arrive at the final estimation sample, we implement the procedure described in \Cref{sec:coursecutoffs} to select applicants with a (locally) preferred and next-best field of study, and exclude a small subset of 751 applicants (representing 1.50\% of the remaining applicants). The excluded applicants are characterized by a combination of preferred and next-best unclustered fields of study with (1) fewer than 10 observations, (2) perfect compliance, where all applicants above (below) the cutoff are enrolled in the preferred (next-best) field in second year, or (3) exact multicollinearity, where all applicants score either above or below the cutoff. The preferred and next-best field combinations with perfect compliance and exact multicollinearity make the estimation of the IV model in \eqref{eq:2stage_rdd}-\eqref{eq:1stage_rdd} impossible. 

Panel A of \Cref{tbl:DescrStats} shows descriptive statistics of the final estimation sample. This sample includes 27.24 percent of the total number of applicants in the four cohorts. The mean of the running variable is positive, and 68\% of the students are assigned to their preferred field. The applicants have on average 5.46 courses on their course ranking. In 88\% of these rankings, the top two fields equal our measure for the preferred and next-best field. The enrolled field in year two is different from the initially matched field for 8\% of the students. The mean of second-year ECTS across all enrolled students is equal to 54.19. However, there is substantial variation in the number of ECTS between fields. For instance, students enrolled in Physical sciences and Architecture and building have a mean second-year ECTS of less than 50, those in Law and Social and behavioural science have a mean of around 53, and students in Health and Journalism and information have a mean of almost 60. 

Panel B of \Cref{tbl:DescrStats} provides an overview of the most common preferred fields for each next-best. All 22 next-best fields have a skewed distribution of students across preferred fields. For instance, 77.83\% (16.64\%) of the students with Computing as next-best prefer Engineering (Business and administration). The remaining 5.53\% prefer any of the other 19 fields. Panel B also shows that the number of observations differs across next-best fields. For example, there are 7,432 students with Business and administration as next-best versus 363 students with Agriculture, forestry and fishery.

\subsection{Robustness checks and validity of the continuity assumption}\label{A:robustness}

We assess the robustness of our results against more flexible specifications. In our baseline specification, we use a polynomial of degree one, so that $f_h(S)=\gamma S$, and do not apply a bandwidth. First, we include an additional interaction term between $S$ and an indicator for scoring above the cutoff across all preferred fields ($\sum_{c \neq b} \tilde{Z}_c$) in \eqref{eq:1stage_rdd}. Second, we include $S^2$ as an additional control variable in both \eqref{eq:2stage_rdd} and \eqref{eq:1stage_rdd}. Third, we assess the robustness of our results via the use of symmetric bandwidths equal to 40, 30, and 20. These bandwidths include 94.59\%, 87.02\%, and 70.94\% of the total sample respectively. 

\begin{figure}[tb!]
\centering
\caption{Robustness checks}
\label{fig:fsrobustness} 
\includegraphics[width=1.0\textwidth]{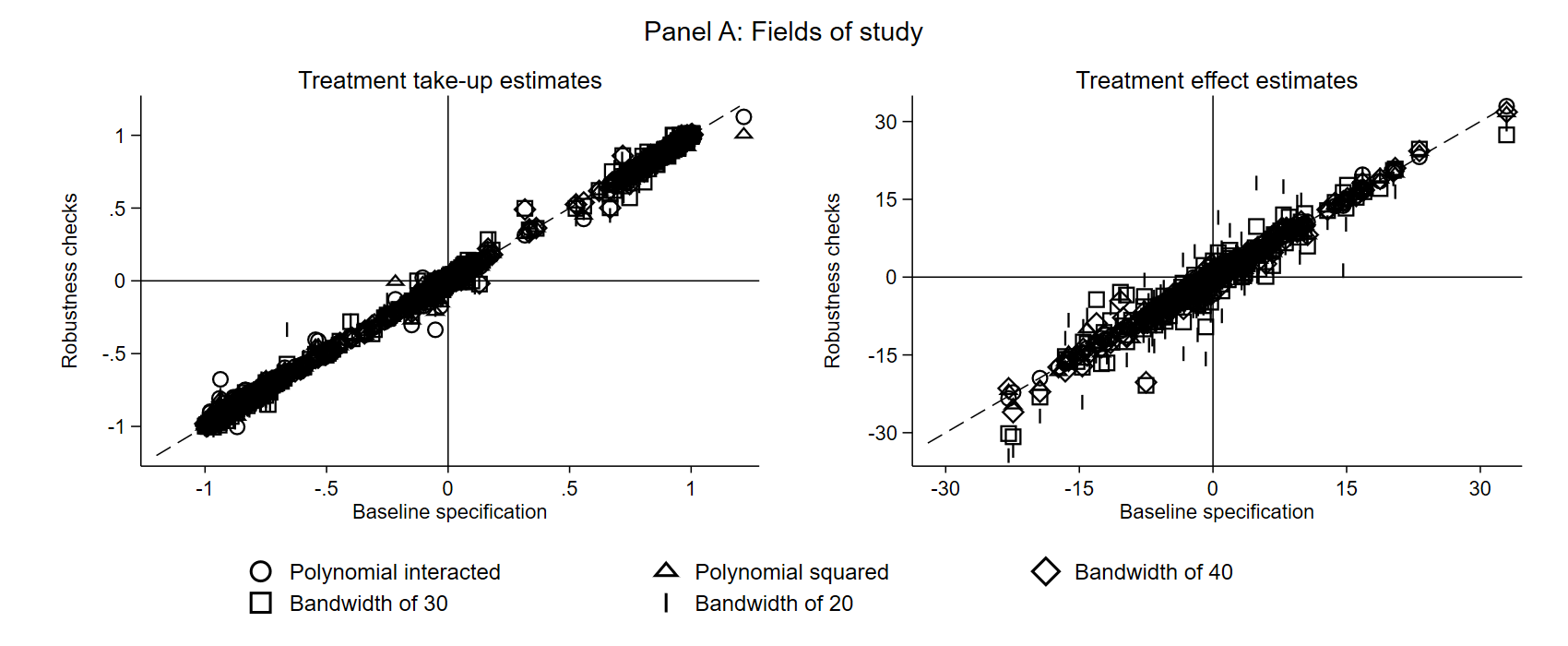} 
\vspace{-2.2cm} \\
\includegraphics[width=1.0\textwidth]{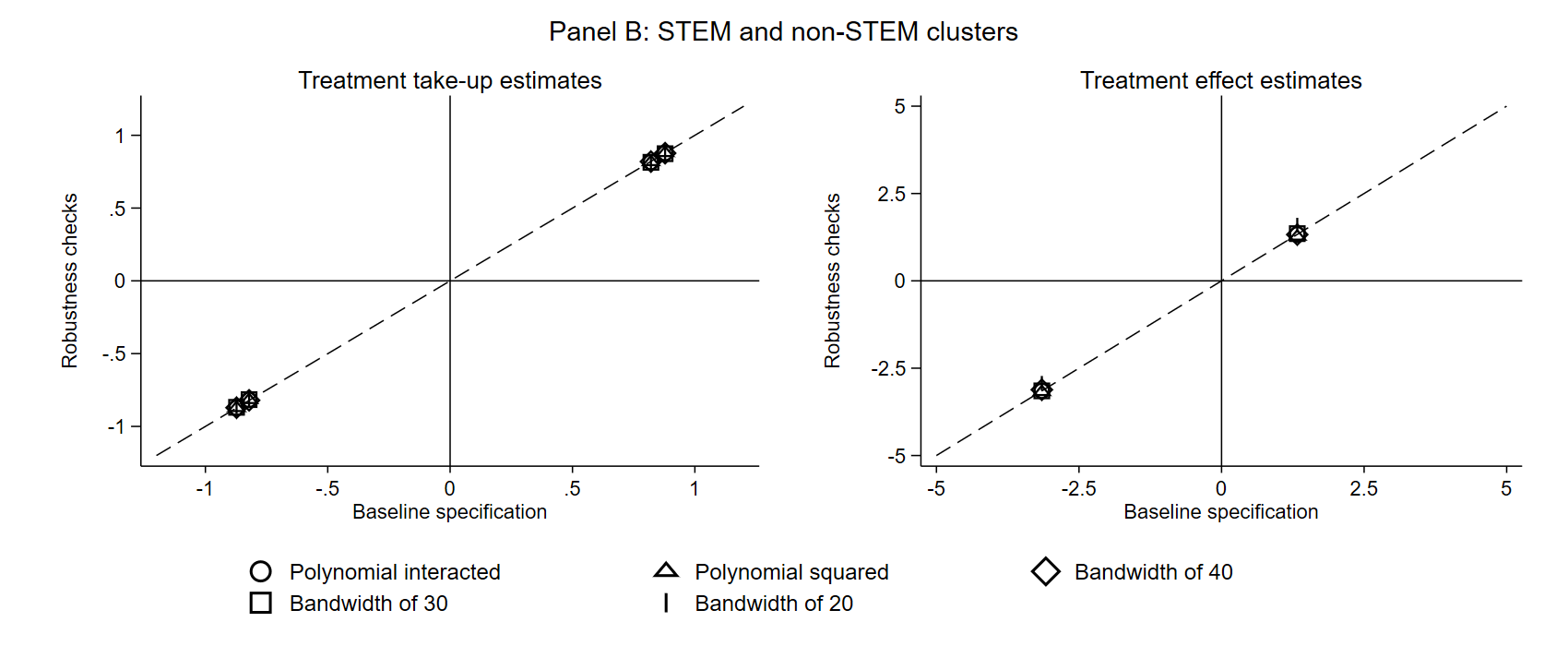}
\vspace{-2.2cm} \\
\includegraphics[width=1.0\textwidth]{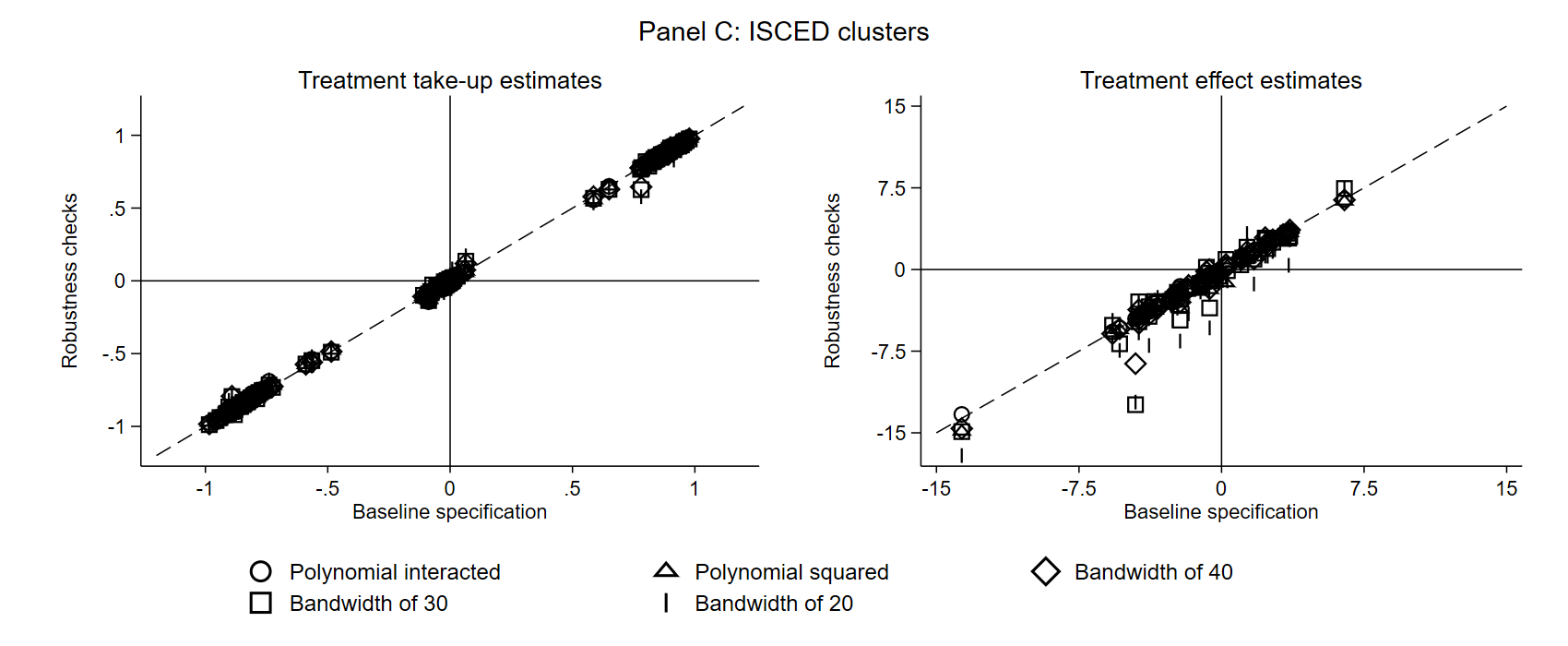}
\caption*{ \footnotesize
Notes: This figure shows scatterplots of the estimates from five robustness checks against those from the baseline specification. The left plots display treatment take-up estimates $\pi_{acb}^h$ from \eqref{eq:1stage_rdd}, while the right plots show treatment effect estimates $\beta_{ab}^h$ from \eqref{eq:2stage_rdd}. Panel A shows the results for the unclustered fields of study, panel B for the STEM classification, and panel C for the ISCED classification.
}
\end{figure}

Panel A of \Cref{fig:fsrobustness} presents the results of these robustness checks using the unclustered fields of study. Specifically, the left plot compares the treatment take-up estimates from \eqref{eq:1stage_rdd} of our baseline specification with those from each robustness check. The estimates lie close to the 45-degree line, indicating virtually identical results across all specifications. Similarly, the right plot illustrates that the treatment effect estimates from \eqref{eq:2stage_rdd}  are similar across the robustness checks. The largest deviation occurs with a bandwidth of 20, which also substantially reduces statistical power and weakens the significance. Panels B and C of \Cref{fig:fsrobustness} confirm that these conclusions continue to hold when using the STEM and the ISCED classification.

\begin{figure}[tb!]
\centering
\caption{Balancing test}
\label{fig:balancingtest} 
\includegraphics[width=.8\textwidth]{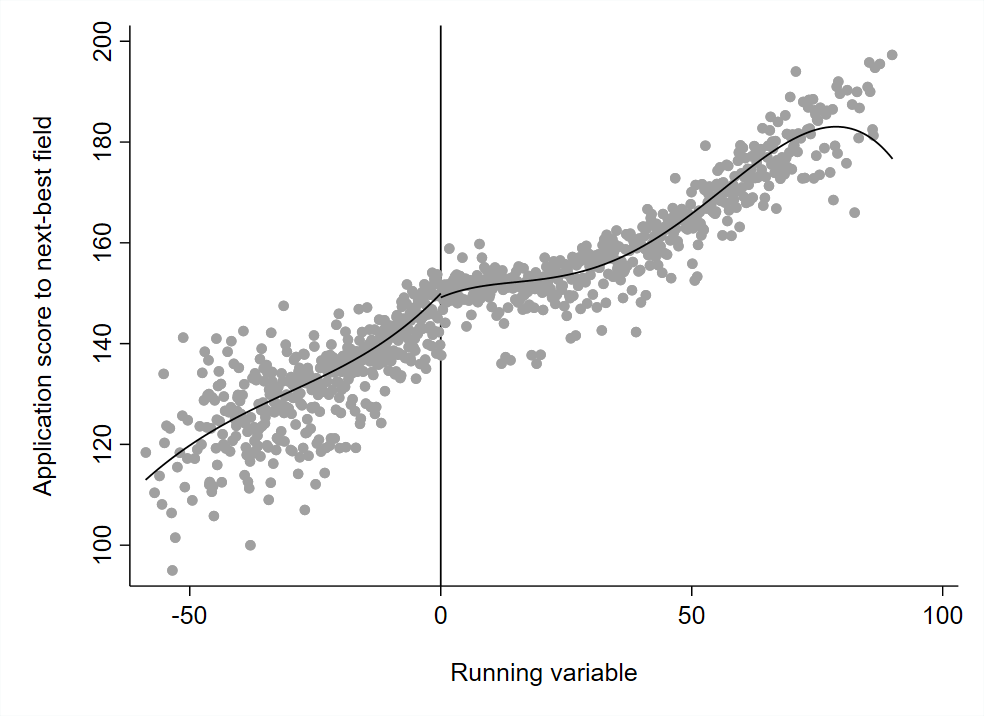} 
\caption*{ \footnotesize
Notes: This figure shows a scatterplot of the application score to the next-best field against the running variable with a fourth-order polynomial estimated with a uniform kernel on the full estimation sample (the default of the Stata program rdplot).
}
\end{figure}

Next, we assess the validity of the local continuity assumption (Assumption~\ref{ass:RD}). First, we examine whether the application score to the next-best field is continuous at the cutoff. \Cref{fig:balancingtest} plots the maximum score across all courses in the preference block of the next-best field against the running variable. Consistent with continuous potential outcomes, this score does not show a discontinuity around the cutoff. This result is confirmed with a pooled regression on the students from all next-best fields: We regress the application score to the next-best field on the running variable and a single treatment dummy indicating whether students scored above the cutoff, and the estimate on this treatment dummy is equal to -0.31 and has a $p$-value of 0.17. The application score is the only background characteristic we observe for all students.

\begin{figure}[tb!]
\centering
\caption{Discontinuity in density test}
\label{fig:discontinuitytest} 
\includegraphics[width=.8\textwidth]{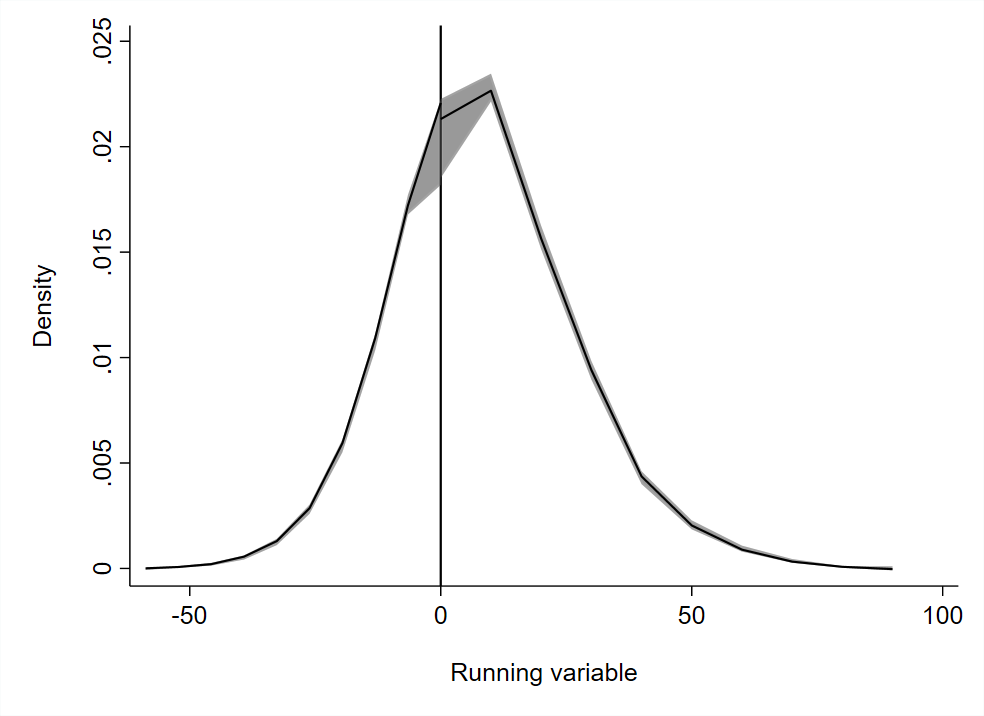} 
\caption*{ \footnotesize
Notes: This figure shows the density of the running variable estimated with a second-order polynomial and a third-order polynomial for the bias-correction estimate, an optimally-chosen bandwidth with the full estimation sample as plot range, a triangular kernel, and 95\% confidence intervals with jackknifed standard errors (the default of the Stata program rddensity).
}
\end{figure}

Second, we examine the density of the students across all next-best fields near the course cutoffs \citep{mccrary2008}. If applicants have precise control over their application score, we might observe bunching just above the cutoffs and the local continuity assumption may not be valid \citep*{lee2010regression}. However, \Cref{fig:discontinuitytest} shows no sign of excess mass just above the cutoffs. The bias corrected discontinuity test introduced in \cite*{CJA2018} has a test statistic equal to 0.13 with a $p$-value of 0.90, implying that we cannot reject the null hypothesis of no discontinuity. 

The continuity assumption is also consistent with the setting at hand. The number of applications is different each year, and hence the cutoffs change from year to year and are unknown at the time of application. The cutoffs only become publicly available when the applicants receive their offer. A regression of the cutoffs across years on course dummies has an R-squared of 0.82, which confirms the movement in the cutoffs across years. 

\subsection{Testing the necessary conditions}
\label{A:testingnecessary}

The (in)equalities $\phi^A_{ljk} \lesseqgtr 0$ in \Cref{prop:nec_cond} can be written as conditional moment inequalities with $Y$ as the conditioning variable instead of $Z$. In particular, the inequality $\phi^A_{ljk} \geq 0$ with $l,j \in S_a$ and $k \in S_b$ can be written as
\begin{align} \label{eqn:ineq1}
   \mathbb{E}[\mathbb{I}_{Y\in A} D_l|Z=k] &\leq \mathbb{E}[\mathbb{I}_{Y\in A} D_l|Z=j] \\ \notag
   \mathbb{E}[\mathbb{I}_{Y\in A} D_l Z_k]\mathbb{P}[Z=j] &\leq 
   \mathbb{E}[\mathbb{I}_{Y\in A} D_l Z_j]\mathbb{P}[Z=k] \\ \notag
   \mathbb{E}[c_j D_l Z_k -& c_k D_l Z_j |Y=y] \leq 0 ,
\end{align}
where $c_j=\mathbb{P}[Z=j]$, and the last line follows from \cite{mourifie2017testing}. Similarly, the inequality $\phi^A_{ljk} \leq 0$ with $l,k \in S_b$ and $j \in S_a$ can be written as
\begin{align} \label{eqn:ineq2}
   \mathbb{E}[\mathbb{I}_{Y\in A} D_l|Z=j] &\leq \mathbb{E}[\mathbb{I}_{Y\in A} D_l|Z=k] \\ \notag
   \mathbb{E}[c_k D_l Z_j -& c_j D_l Z_k |Y=y] \leq 0 .
\end{align}
For the equality $\phi^A_{ljk}=0$ with $l\notin \{S_a,S_b\}$, $j \in S_a$, and $k \in S_b$, both \eqref{eqn:ineq1} and \eqref{eqn:ineq2} must hold.

These conditional moment inequalities can be tested using the intersection bounds framework of \cite{chernozhukov2013intersection}, implemented in the Stata package \textit{clrtest} by \cite{chernozhukov2015implementing}. The \textit{clrtest} allows for the joint testing of inequalities within an estimation sample. Hence, within each sample defined by a preferred and next-best field $(j,k)$ at the unclustered fields of study level, we jointly test the restrictions on $\phi^A_{ljk}$ for all treatments $l$.

For example, similar to the discussion of the identifying assumptions in \Cref{sec:identifyingass}, consider testing the STEM classification with Personal services as the next-best field ($k$) in the non-STEM cluster. In this case there are 8 preferred STEM fields ($j$), such as Physical sciences. An offer from $j$ instead of $k$ can weakly increase enrollment in the 8 STEM fields according to \eqref{eqn:ineq1}, and weakly decrease it in the 14 non-STEM fields according to \eqref{eqn:ineq2}. 

Now consider testing the ISCED classification. Again take the next-best field Personal services ($k$), which now belongs to the Services cluster, and the preferred field Physical sciences ($j$), which belongs to the Science cluster. In this case an offer from $j$ instead of $k$ can weakly increase enrollment in the 4 fields belonging to the Science cluster according to \eqref{eqn:ineq1}, weakly decrease it in the 4 fields in the Services cluster according to \eqref{eqn:ineq2}, and should have no effect on the 14 fields in the other clusters according to \eqref{eqn:ineq1} and \eqref{eqn:ineq2}.

To account for multiple testing within each next-best field $k$, we follow \cite{mourifie2017testing} and apply the \cite{holm1979simple} correction to multiple applications of \textit{clrtest} across different samples. The Holm procedure orders the $p$-values within each next-best field and compares them to adjusted significance levels $\alpha$ divided by their rank. Rejections stop once a $p$-value exceeds its threshold. Because \textit{clrtest} does not report $p$-values directly, we retrieve them by evaluating the test over a fine grid of confidence intervals in $[0.90,1)$, constructed with step sizes that decrease from $10^{-3}$ to $10^{-12}$ as the grid approaches 1.

The \textit{clrtest} package allows both nonparametric and parametric estimation of the conditional moment inequalities. We use the parametric option, which approximates the conditional expectations using a linear regression in $Y$. This is preferable in our setting because the estimation samples $(j,k)$ can be small, making nonparametric estimation less reliable.

The inequalities are evaluated over values of the conditioning variable $Y$. In our application $Y$ corresponds to the number of ECTS credits. We evaluate the inequalities for values of $Y$ between the 20th percentile (48 ECTS) and the 80th percentile (60 ECTS), in increments of 1 ECTS, avoiding extreme outcome values.

We implement the tests without including the running variable. Since the inequalities are intended to hold locally at the application score cutoff, this corresponds to the local randomization approach to RDDs \citep{cattaneo2024elements}. In the baseline test we use the full sample, as in the main RDD analysis. As a robustness check we also repeat the tests using symmetric bandwidths of 40, 30, and 20. The results are similar.

\subsection{Treatment take-up estimates with ISCED classification}
\label{A:takeup}

\begin{table}[t!]
    \begin{adjustwidth}{-1.45cm}{-1.45cm}
    \caption{Treatment take-up estimates with ISCED clusters}
    \label{tab:isced_fs}
    \renewcommand{\arraystretch}{0.75}
    \begin{tabular}{lcccccccc}
    \toprule \toprule
Instrument $\widetilde{Z}_a$  & \multicolumn{8}{c}{Treatment cluster $S_b$}\\
\cmidrule(lr){1-1}\cmidrule(lr){2-9}
   & Edu- & Hum. & Soc. sci., & Science & Eng., & Agri- & Health & Ser- \\ [-.5em]
& cation & and& bus. and& & man. and & culture& and & vices\\  [-.5em]
&  & arts& law& & constr. & & welfare & \\ 
\midrule
Education	&		&	0.84***	&	0.94***	&		&		&		&	0.95***	&	0.98***	\\ [-.5em]
	&		&	(0.06)	&	(0.02)	&		&		&		&	(0.03)	&	(0.02)	\\
Humanities and arts	&	0.86***	&		&	0.86***	&	0.85***	&	0.82***	&		&	0.95***	&	0.81***	\\ [-.5em]
	&	(0.05)	&		&	(0.01)	&	(0.06)	&	(0.03)	&		&	(0.03)	&	(0.04)	\\
Soc. sci., bus. and law	&	0.87***	&	0.83***	&		&	0.83***	&	0.86***	&		&	0.93***	&	0.90***	\\ [-.5em]
	&	(0.02)	&	(0.01)	&		&	(0.02)	&	(0.02)	&		&	(0.01)	&	(0.02)	\\
Science	&		&	0.78***	&	0.88***	&		&	0.87***	&	0.89***	&	0.89***	&	0.88***	\\ [-.5em]
	&		&	(0.10)	&	(0.02)	&		&	(0.01)	&	(0.07)	&	(0.01)	&	(0.03)	\\
Eng., man. and constr.	&		&	0.86***	&	0.91***	&	0.84***	&		&	0.90***	&	0.87***	&	0.65***	\\ [-.5em]
	&		&	(0.03)	&	(0.01)	&	(0.01)	&		&	(0.05)	&	(0.02)	&	(0.04)	\\
Agriculture	&		&		&	0.96***	&	0.78***	&	0.86***	&		&	0.83***	&	0.78***	\\ [-.5em]
	&		&		&	(0.04)	&	(0.04)	&	(0.05)	&		&	(0.03)	&	(0.13)	\\
Health and welfare	&	0.91***	&	0.94***	&	0.90***	&	0.81***	&	0.83***	&	0.59***	&		&	0.92***	\\ [-.5em]
	&	(0.03)	&	(0.03)	&	(0.01)	&	(0.02)	&	(0.02)	&	(0.04)	&		&	(0.02)	\\
Services	&	0.96***	&	0.86***	&	0.90***	&	0.91***	&	0.88***	&	0.80***	&	0.92***	&		\\ [-.5em]
	&	(0.03)	&	(0.03)	&	(0.01)	&	(0.03)	&	(0.03)	&	(0.10)	&	(0.03)	&		\\
Observations   & 1370 & 4610 & 17138 & 9693 & 8379 & 1079 & 4119 & 2884\\ 
      \bottomrule \bottomrule
    \end{tabular}
    \begin{tablenotes}
    \footnotesize
    \item[] Notes: This table reports estimates of $\pi_{acb}^h$ from \eqref{eq:1stage_rdd} for enrollment $\widetilde{D}_a$ in the cluster $S_a$ containing the preferred field, using the ISCED classification. The columns correspond to treatment clusters $S_b$ and the rows to treatment clusters $S_a$. The final row reports the size of the estimation sample comprising students whose next-best field belongs to the corresponding cluster $S_b$. The estimates are obtained using OLS, with robust standard errors in parentheses. Empty off-diagonal entries indicate pairs of treatment clusters for which all combinations of preferred and next-best fields were excluded because of insufficient observations (see \Cref{A:data} for details). Stars indicate statistical significance at the ${}^{*}10$, ${}^{**}5$, and ${}^{***}1$\% levels after applying a Holm correction for the number of tests within each estimation sample.
    \end{tablenotes}
\end{adjustwidth}
\end{table}

We analyze the treatment take-up estimates for the ISCED classification. Estimating \eqref{eq:1stage_rdd} yields a matrix of treatment take-up effects for each treatment cluster $S_b$. For each combination of clusters $S_a$ and $S_b$ containing the preferred and next-best fields, respectively, \Cref{tab:isced_fs} reports only the effect of receiving an offer from the preferred field on enrollment in a field from cluster $S_a$. Most estimates are large and precise, ranging from 0.80 to 0.90. Thus, scoring above the cutoff and receiving an offer from the preferred field raises the probability of enrollment in the cluster containing that field by 80 to 90 percentage points. The smallest estimate is 0.59 for students preferring a field in Health and welfare over one in Agriculture. Overall, these results provide strong evidence that Assumption~\ref{ass:relevance} holds for the ISCED classification.

\subsection{Additional empirical results}
\label{A:additionalresults}

This section applies the results derived in Appendix~\ref{A:Weights} to the field of study level data to provide additional information about the weights and LATEs underlying the estimated treatment cluster effects. The required treatment take-up estimates $\pi_{ljk}^h$ from \eqref{eq:1stage_rdd} and treatment effect estimates  $\beta_{jk}^h$ from \eqref{eq:2stage_rdd} at the unclustered field of study level are provided as supplementary data in the Supporting Information. Because we do not conduct statistical inference on the estimated bounds or weighted averages, we use them only to describe patterns in our sample and draw no conclusions about their population-level counterparts.

\subsubsection{Two treatment clusters: STEM and non-STEM}

\begin{figure}[tb!]
\centering
\caption{Treatment effect weights and weighted LATEs within STEM clusters}
\label{fig:stem} 
\includegraphics[width=1.0\textwidth]{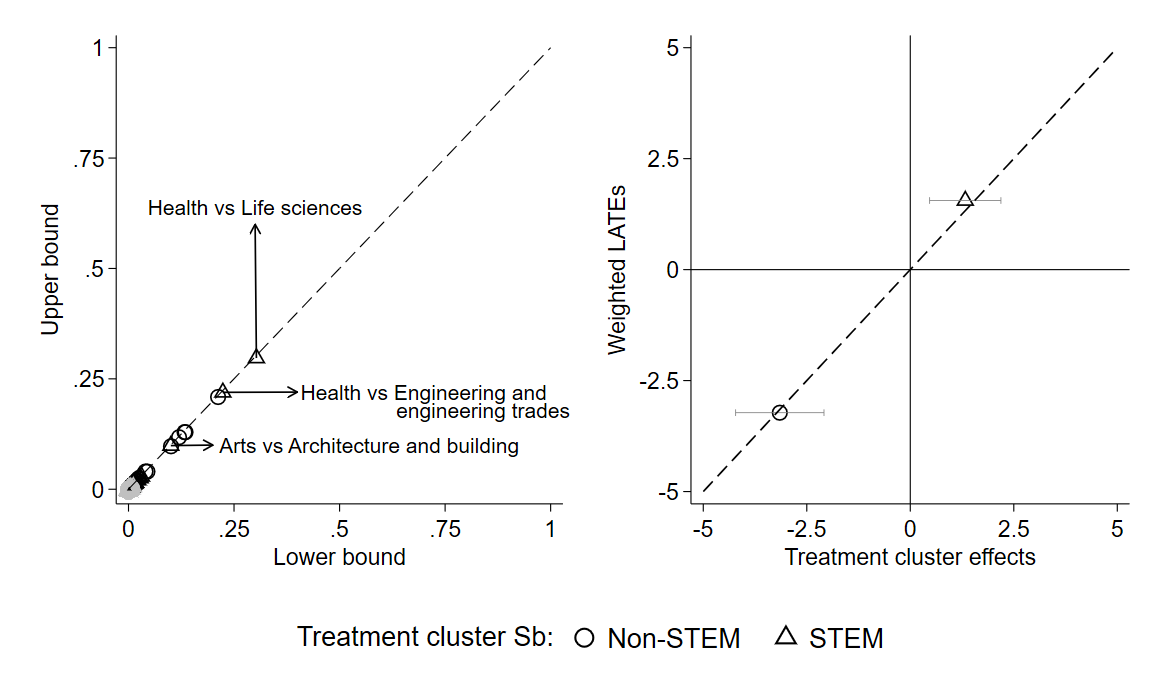} 
\caption*{ \footnotesize
Notes: The left panel shows a scatterplot of the estimated upper bounds of the treatment effect weights against the estimated lower bounds, following \eqref{eq:weights} from \Cref{cor:weights}, for both treatment clusters $S_b$ using the STEM classification. The black (gray) markers correspond to the weights $\lambda_{jklm}$ with $j=l$ and $k=m$ ($j \neq l$ and/or $k \neq m$). The right panel shows a scatterplot of the estimated weighted average of LATEs from \eqref{eq:wa_beta} in \Cref{cor:homo} against the estimated treatment cluster effects, for both treatment clusters $S_b$ using the STEM classification. The lines are 95\% confidence intervals for the treatment cluster effects.
}
\end{figure}

We estimate the bounds on the weights in \Cref{cor:weights} by replacing the treatment take-up estimands $\pi_{ljk}$ and the relevant probabilities with their sample analogues. This yields estimated bounds for the 12,544 treatment effect weights between the STEM and non-STEM clusters. The left plot in \Cref{fig:stem} shows a scatterplot of the estimated upper bounds against the estimated lower bounds for the weights $\lambda_{jklm}$. The black (gray) markers correspond to the weights with $j=l$ and $k=m$ ($j \neq l$ and/or $k \neq m$). The estimated bounds are tight and suggest that most weight is assigned to response types $(D(j)=j,D(k)=k)$ with $j \in S_a$ and $k \in S_b$. For instance, the treatment cluster effect estimate of non-STEM versus STEM consists for 30\% of the LATE of Health versus Life sciences with offers from respectively Health and Life sciences, another 22\% is attributed to Health versus Engineering and engineering trades with offers from respectively Health and Engineering and engineering trades, and 10\% to Arts versus Architecture and building with offers from respectively Arts and Architecture and building. 

\begin{figure}[tb!]
\centering
\caption{Weighted LATEs within STEM and ISCED clusters and response type restrictions}
\label{fig:stem_isced} 
\includegraphics[width=1.0\textwidth]{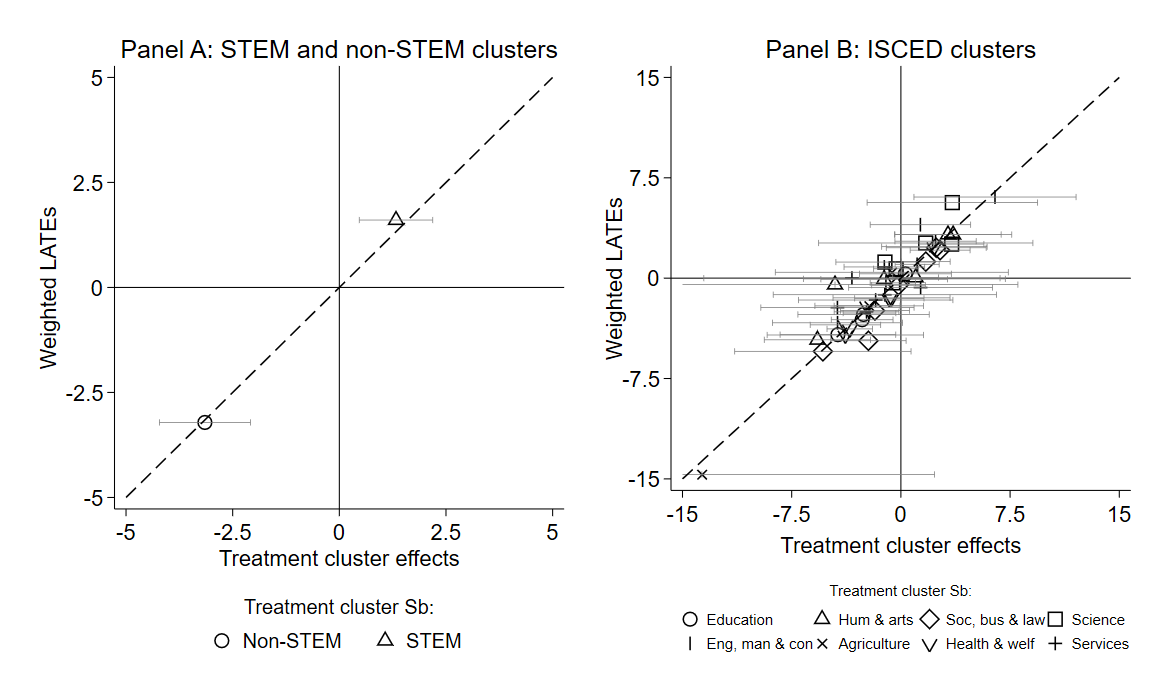} 
\caption*{ \footnotesize
Notes: The left panel shows a scatterplot of the estimated weighted average of LATEs from \eqref{eq:singleshift} in \Cref{cor:alpha} against the estimated treatment cluster effects, for both treatment clusters $S_b$ using the STEM classification. The lines are 95\% confidence intervals for the treatment cluster effects. The right panel shows the same scatterplot for all treatment clusters $S_b$ using the ISCED classification.
}
\end{figure}

% lates in clate - shift assumptions
We next examine the assumptions that allow for the identification of the LATEs within the weighted averages. Corollaries \ref{cor:homo} and \ref{cor:alpha} show that the LATEs are identified by $\alpha_{jk}$ in \eqref{eq:alpha} under treatment effect heterogeneity restrictions or additional response type restrictions, respectively. Under either set of restrictions, the treatment cluster effects between STEM and non-STEM equal a weighted average of identifiable LATEs. We estimate these weighted averages by replacing the treatment take-up and LATE estimands, $\pi_{ljk}$ and $\alpha_{jk}$, and the relevant probabilities with their sample analogues. The right plot in \Cref{fig:stem} plots the estimated weighted average under treatment effect heterogeneity restrictions from \Cref{cor:homo} against the estimated treatment cluster effects with 95\% confidence intervals. The left plot of \Cref{fig:stem_isced} shows these estimates under response type restrictions  from \Cref{cor:alpha}. In both figures, the confidence intervals overlap with the 45-degree line. One explanation for this result may be that treatment effect heterogeneity is restricted. However, it may be more likely that the weighted averages of LATEs assign little weight to the response types $(D(j)=l, D(k)=m)$ with $j \neq l$ and/or $k \neq m$. 

The necessary conditions and the weighted average of LATEs suggest that the estimates according to \eqref{eq:alpha} provide a reasonable approximation to the LATEs in the treatment cluster effects. Hence, we can further unpack the treatment cluster effects. For instance, the non-STEM versus STEM treatment cluster effect (1.33 credits) comprises 30\% from Health versus Life sciences (-0.03 credits), 22\% from Health versus Engineering and engineering trades (1.08 credits), and 10\% from Arts versus Architecture and building (5.02 credits).
With an average absolute value of the estimated LATEs of $4,628$ ECTS credits, all ten estimates with Environmental protection as next-best field are considered unreliable and set to zero when constructing the weighted average of LATEs in Corollaries \ref{cor:homo} and \ref{cor:alpha}. 

\subsubsection{Multiple treatment clusters: Eight ISCED clusters}

\begin{figure}[tb!]
\centering
\caption{Treatment effect weights and weighted LATEs within ISCED clusters}
\label{fig:isced} 
\includegraphics[width=1.0\textwidth]{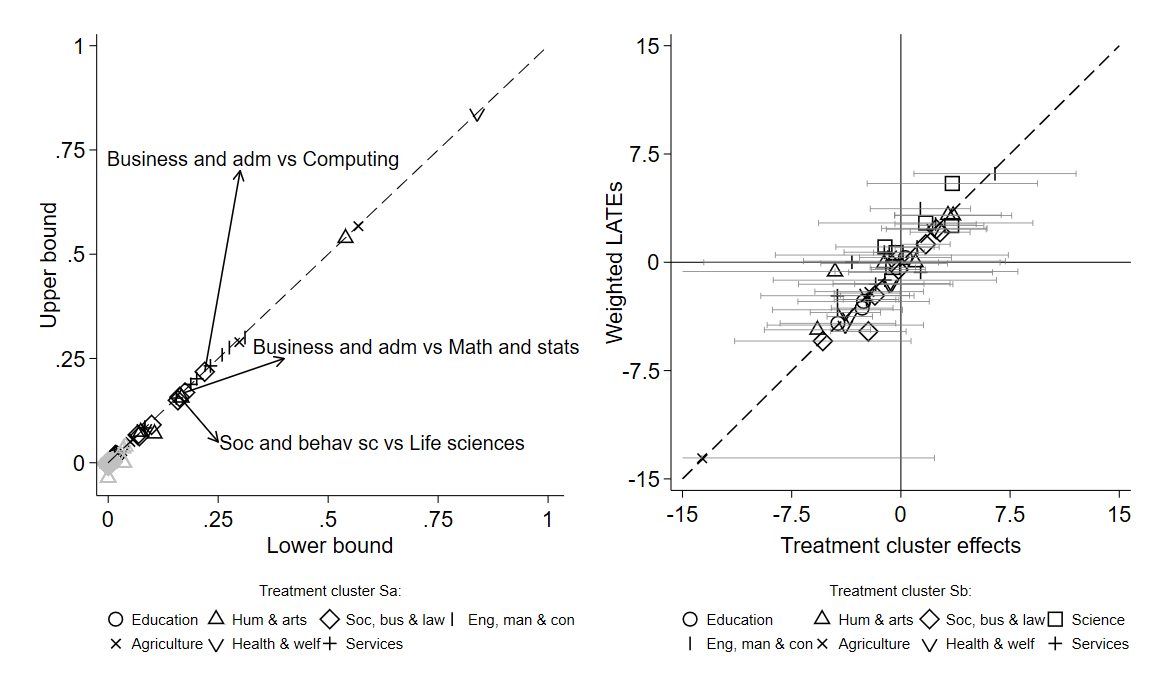} 
\caption*{ \footnotesize
Notes: The left panel shows a scatterplot of the estimated upper bounds of the treatment effect weights against the estimated lower bounds, following \eqref{eq:weights} from \Cref{cor:weights}, for each treatment cluster $S_a$ containing the preferred field when the next-best field belongs to the Science cluster $S_b$, using the ISCED classification. The black (gray) markers correspond to the weights $\lambda_{jklm}$ with $j=l$ and $k=m$ ($j \neq l$ and/or $k \neq m$). The right panel shows a scatterplot of the estimated weighted average of LATEs from \eqref{eq:wa_beta} in \Cref{cor:homo} against the estimated treatment cluster effects, for all treatment clusters $S_b$ using the ISCED classification. The lines are 95\% confidence intervals for the treatment cluster effects.
}
\end{figure}

% weights
We bound the weights for the treatment effects between the eight clusters. The left plot in \Cref{fig:isced} shows the estimated weights when the next-best field belongs to the Science cluster, allowing us to distinguish between the seven clusters containing the preferred fields. As with the STEM clusters, the estimated bounds are tight and suggest that most weight is assigned to response types $(D(j)=j,D(k)=k)$ with $j \in S_a$ and $k \in S_b$. For instance, the treatment cluster effect for Social sciences, business and law versus Science consists for 22\% of the LATE of Business and administration versus Computing with offers from respectively Business and administration and Computing. Another 17\% is attributed to Business and administration versus Mathematics and statistics with offers from Business and administration and Mathematics and statistics, and 16\% to Social and behavioural science versus Life sciences with offers from Social and behavioural science and Life sciences. 

We examine the additional assumptions to identify the LATEs within the treatment cluster effects. The right plot in \Cref{fig:isced} plots the estimated weighted average of LATEs against the treatment effects between the ISCED clusters using treatment effect heterogeneity restrictions. The right plot of \Cref{fig:stem_isced} shows these estimates under the response type restrictions. Similar to the results with the STEM clusters, the 95\% confidence intervals of the treatment cluster effects overlap with the 45-degree line. This may indicate negligible bias in the identification of the LATEs due to little treatment effect heterogeneity or little weight to the response types $(D(j)=l, D(k)=m)$ with $j \neq l$ and/or $k \neq m$.

% lates
In this case, the estimated LATEs corresponding to the 22 fields of study can be used to provide further insights into the treatment cluster effects. For instance, the estimated treatment cluster effect of Social sciences, business and law versus Science (3.49 credits) comprises 22\% from Business and administration versus Computing (2.60 credits), 17\% from Business and administration versus Mathematics and statistics (6.69 credits), and 16\% from Social and behavioural science versus Life sciences (4.87 credits).

\end{document}